\documentclass[useAMS,usenatbib]{mn2e}
\usepackage[dvips]{graphicx}
\usepackage{mathrsfs}
\usepackage{amsmath}
\usepackage{times}
\newcommand{\ud}{\mathrm{d}}
\newcommand{\vers}[1]{\boldsymbol{\hat{e}_#1}}
\newcommand{\de}{\partial}
\newcommand{\rot}[1]{\boldsymbol{\nabla} \times \boldsymbol{#1} }
\newcommand{\PP}{\mathscr{P}}
\newcommand{\B}{\boldsymbol{B}}
\newcommand{\nnabla}{\boldsymbol{\nabla}}

\title[Topology of magnetars external field. I.]{Topology of magnetars external field.   I. Axially symmetric fields}
\author[Pavan, L., Turolla, R., Zane, S. and Nobili, L.]{L. Pavan$^{1}$, R. Turolla$^{1, 2, 3}$,
S. Zane$^{2}$ and L. Nobili$^{1}$\\
$^{1}$Department of Physics, University of Padova,  via Marzolo 8, 35131 Padova, Italy\\
$^{2}$Mullard Space Science Laboratory, University College London, Holmbury St. Mary, Dorking, Surrey, RH5 6NT, UK\\
$^{3}$INFN, Sezione di Padova,  via Marzolo 8, 35131 Padova, Italy }
\begin{document}
\date{Accepted ... Received ...}
\pagerange{\pageref{firstpage}--\pageref{lastpage}} \pubyear{}
\maketitle

\label{firstpage}

\begin{abstract}

There is an increasing theoretical and observational evidence that the external magnetic field of magnetars may
contain a toroidal component, likely of the same order of the poloidal one. Such ``twisted magnetospheres'' are
threaded by currents flowing along the closed field lines which can efficiently interact with soft thermal photons via
resonant cyclotron scatterings (RCS). Actually, RCS spectral models proved quite successful in explaining the
persistent $\sim 1$--10 keV emission from the magnetar candidates, the soft $\gamma$-ray repeaters (SGRs) and
the anomalous X-ray pulsars (AXPs). Moreover, it has been proposed that, in presence of highly relativistic
electrons, the same process can give rise to the observed hard X-ray spectral tails extending up to $\sim 200$
keV.

Spectral calculations have been restricted up to now to the case of a globally twisted dipolar magnetosphere, although
there are indications that the twist may be confined only to a portion of the
magnetosphere, and/or that the large scale field is more complex than a simple dipole. In this paper we
investigate multipolar, force-free magnetospheres of ultra-magnetized neutron stars. We first discuss  a general
method to generate multipolar solutions of the Grad-Schl\"uter-Shafranov equation, and analyze in detail
dipolar, quadrupolar and octupolar fields. The spectra and lightcurves for these multipolar, globally twisted
fields are then computed using a Monte Carlo code  and compared with those of a purely dipolar configuration.
Finally the phase-resolved spectra and energy-dependent lightcurves obtained with a simple model of a locally
sheared field are confronted with the {\em INTEGRAL} observations of the AXPs 1RXS J1708-4009 and  4U 0142+61.
Results support a picture in which the field in these two sources is not globally twisted.
\end{abstract}

\section{Introduction}\label{intro}
Among isolated neutron stars (INSs), a small number of objects, namely the Soft Gamma Repeaters (SGRs) and the
Anomalous X-ray Pulsars (AXPs), share a number of common, peculiar characteristics. These are the huge spin down
rates, $\dot P\sim 10^{-10}$--$10^{-13} \ {\rm s/s}$, absence of radio emission \citep[up to now only the AXPs
XTE J1810-197 and 1E 1547.0-5408 have been detected in the radio, ][]{camilo1, camilo2}, persistent X-ray luminosities
in the range $L\sim 10^{34}$--$10^{36}$ erg/s and  rotational periods $P\sim 2$--12 s, a very narrow
range compared with that of other classes of INSs. Though SGRs activity is higher, both groups undergo erratic
X/$\gamma$-ray bursts with peak luminosities of $\sim 10^{40}$--$10^{41}$ erg/s and typical durations of $\sim
0.1$--1 s. Moreover, SGRs exhibit much more energetic events, the so-called Giant Flares (GFs), in which
energies up to $\sim 10^{47}$ erg are released in a timescale of one second \citep[see, for a recent
review,][]{mereghetti}.
Thanks to {\em INTEGRAL} ISGRI and {\em RXTE} HEXTE, AXPs and SGRs are now known to be also
persistent hard X-ray sources \cite[e.g.][]{khm04,k06,m05,gotz}. Actually, their energy output in the
$\sim 20$--200 keV range may amount to as much as 50\% of the total flux emitted above $\sim 1$ keV.
Recent, deeper {\em INTEGRAL} observations have  shown
that the hard X-ray emission is highly phase-dependent and probably results from the
superposition of different spectral components \citep{dhartog1, dhartog2}.

At variance with radio-pulsars, the persistent emission of SGRs/AXPs is $\sim 10$--100 times higher than their
rotational energy losses. This, together with the lack of detected stellar companions,
indicates that the persistent emission of these sources is unlikely to be powered by rotation or accretion. The
(dipolar) magnetic fields inferred from the spin-down measurements, $B\sim 10^{14}-10^{15}$ G, largely in excess
of the quantum critical field ($B_Q=4.4\times 10^{13}$ G), support the idea that SGRs and AXPs are
ultra-magnetized NSs \citep[or magnetars;][]{dt92, td93} and their (persistent and bursting) emission is
sustained by the super-strong magnetic field. Although other scenarios,
mainly based on accretion
from a debris disc left after the supernova event \citep[e.g.][]{disk1, disk2, disk3}, may still be considered,
the magnetar model appears capable of explaining in a simple and economical way most of the observed properties
of SGRs and AXPs \citep[e.g.][]{wt06}.

In a magnetar with surface field $\sim 10^{15}$ G the internal field can reach  $\sim 10^{17}$ G
\citep{td93, td95}. Since the poloidal and toroidal components are expected to be in rough equipartition
\citep[see][and references therein]{tlk}, the huge toroidal field stresses the crust, producing a
deformation of the surface layers. This, in turns, induces a rotation of the external field lines which are
anchored to the star crust and leads to the appearance of an external toroidal component. The properties of such
a twisted magnetosphere have been investigated by \cite{tlk} by means of a model analogous to that for the
solar magnetic field \citep[e.g.][]{low+lou, wolfson}, under the assumptions of a static, dipolar,
globally twisted field and enforcing, as in the solar models, the force-free condition.

A feature of (non-potential)
force-free fields is the presence of supporting currents. As first
suggested by \cite{tlk}, thermal photons emitted by the star surface can
scatter at the cyclotron resonance on the charges flowing in the
magnetosphere, and this can drastically alter the primary spectrum.
Recent, detailed calculations \citep{lg06, fern, ntz1} of scattering onto
mildly relativistic electrons confirmed this
picture. Typical synthetic spectra exhibit a high-energy tail,
superimposed to a thermal bump and closely resemble
the (empirical) ``blackbody+power-law'' model which has been routinely used to
describe the magnetars quiescent emission in the
$\sim 0.5$--10 keV band \cite[see again][for a summary of observational
results]{wt06, mereghetti}. This model has been
successfully applied to the X-ray spectra
of several AXPs/SGRs by \citet[see also
\citealt{ntz1}]{rea}, providing direct support to the twisted
magnetosphere scenario.

The origin of the
high-energy tails discovered
with {\em INTEGRAL} is much less understood. \cite{tb05}
analyzed different mechanisms within the magnetar model, and
suggested that the hard X-rays may be
produced either by thermal bremsstrahlung in the surface layers heated by
returning currents, or by synchrotron emission from pairs created higher
up in the magnetosphere. Quite interestingly, \cite{bh07} and
\cite{bh08} have recently proposed a further possibility, according to
which the soft gamma-rays may also originate from resonant up-scattering
of seed photons,  if a population of highly relativistic electrons is
present in the magnetosphere \cite[see
also][]{ntz2}.

Since the conduction current in a sheared
magnetosphere is $\propto \nnabla \times \B$, the particle density and
hence the optical depth to resonant scattering depends on the shear
\citep{tlk}. Moreover, the global topology of the magnetic field
influences scattering even for fixed shear. A globally twisted dipolar
field will in general give rise to a different spectrum from that of a
globally twisted quadrupolar field with the same twist angle. Because the
spatial distribution of charges is not homogeneous \citep{tlk}, changes in
shear and/or field topology are going to produce differences not only in
the spectral shape, but also in the pulse shape of the received radiation.
Some complicated pulse profiles from SGRs have already been explained by
invoking higher order multipoles \citep[e.g.][]{td01, feroci}.
Moreover, recent data both for AXPs and SGRs seem to point towards the
presence of a localized, rather than global, twist \citep[e.g.][]{woods2,
perna}. However, no investigation which includes multipolar components or
localized twists have been presented up to now.

In this paper we discuss how globally twisted magnetostatic equilibria can be derived in the case of higher
order, axially symmetric multipolar fields. In particular, explicit (numerical) solutions for quadrupolar and
octupolar fields are presented. We use the Monte Carlo code developed by \cite{ntz1} to investigate the
properties of the emerging spectrum and pulse profiles for higher order multipolar fields. The paper is
organized as follows: in \S \ref{sec:dipol-conf-beyond} we introduce the globally twisted model and derive the
solutions for each axially symmetric multipole; an analytical solution, valid in the case of dipolar and
quadrupolar fields for small shear, is presented in Appendix \ref{sec:app}. Monte Carlo spectra and
lightcurves obtained with different force-free magnetospheric configurations are discussed in \S
\ref{sec:spectra}, where a comparison with the timing properties of the hard X-ray emission from the AXPs
1RXS J1708-4009 and  4U 0142+6 is also presented. Discussion follows in \S \ref{sec:conclusions}.

\section{Globally-twisted axisymmetric models}\label{sec:dipol-conf-beyond}

In  this section we present the basic equations that we use to derive globally-twisted magnetospheric models. We
follow closely the approach outlined in \cite{wolfson} and \cite{tlk} (hereafter W95 and TLK, respectively), who
considered the global twist of a dipolar field. As we show below, the same approach can be used to compute
twisted fields for multipoles of arbitrary order. As in W95 and TLK, we restrict to magnetostatic, force-free
equilibria. In the case of a low density, static plasma, in fact, in the standard MHD equation
\begin{equation*}
  \rho \frac{\de \boldsymbol{v}}{\de t} +
  \rho (\boldsymbol{v} \cdot \nnabla) \boldsymbol{v} =
  -\nabla p +\rho \boldsymbol{g} + \boldsymbol{j}\times \B,
\end{equation*}
where $\rho$ and $ \boldsymbol{v}$ are the
plasma density and velocity, the velocity and gravity terms can be neglected. With the further hypothesis that the
(plasma) pressure force is small with respect to the Lorentz force $\boldsymbol{j}\times \B$ (where
$\boldsymbol{j}$ is the current density), the equation reduces to $\boldsymbol{j}\times \B = 0$. Since we are
interested in stationary configurations, the Amp\'ere-Maxwell equation simplifies to $\nnabla \times \B = (4\pi
/ c) \boldsymbol{j}$. From the two previous conditions the usual expression for the force-free condition is
recovered
\begin{equation}
  \label{fff}
  (\nnabla \times \B) \times \B = 0.
\end{equation}

Our aim is to construct an axisymmetric, force-free field by adding a defined amount of shear to a potential
field. In accordance with both TLK and W95, we choose to use the flux function $\PP$ to express the poloidal
component of the field. Axisymmetry is enforced choosing a function independent of the azimuth $\phi$ (we use
a spherical coordinate system with the polar axis along the magnetic moment vector). Thus, the
general expression for an axisymmetric field is:
\begin{equation}
  \label{eq:4}
  \B = \frac{\nnabla \PP(r,\theta) \times \vers{\phi}}{r \sin\theta}
  + B_\phi(r,\theta) \vers{\phi}.
\end{equation}
where $\vers{\phi}$ is the unit vector in the $\phi$ direction.

By inserting the previous expression into the force-free condition (equation [\ref{fff}])
one obtains two independent scalar equations for the flux function, $\PP$, and the
toroidal component of the field, $B_\phi$. The former requires $B_\phi$ to be a function
of $\PP$ only \citep{low+lou}, thus we can write

\begin{equation}
  B_\phi(r,\theta)=  \frac{1}{r \sin\theta} F(\PP)
\end{equation}
where $F$ is an arbitrary function. Introducing the previous expression into the second
scalar equation leads to the Grad-Schl\"uter-Shafranov (GSS) equation
\begin{equation}
  \label{gss}
  \frac{\de^2\PP}{\de r^2} + \frac{1-\mu^2}{r^2}
  \frac{\de^2\PP}{\de \mu^2}+F(\PP)\frac{\ud F}{\ud \PP} =0
\end{equation}
(here and in the following $\mu\equiv\cos\theta$).

The GSS equation can be reduced to an ordinary differential
equation by making suitable assumptions on the dependence of $\PP$ on the
coordinates. Following a classical approach to this problem
\citep[e.g.][W95, TLK]{low+lou}, we assume separation of variables and
choose the flux function $\PP$ in the
form
\begin{equation}
  \PP=\PP_0  \left(\frac{r}{R_{NS}}\right)^{-p} f(\mu)
  \label{ppdep}
\end{equation}
where $f(\mu)$ is a function of the colatitude $\theta$, $R_{NS}$ is the stellar radius
and $\PP_0 = B_{\textrm{pole}}R^2_{NS}/2$, as in TLK.
The requirement that all the components of $\B$ have the same radial dependence implies
that

\begin{equation}\label{fp}
  F(\PP)=\sqrt{\frac{C}{p(1+p)}}\PP^{1+1/p}
\end{equation}
where, for later convenience, we have expressed the multiplicative constant in terms
of a parameter $C$ and of the radial exponent $p$. Recalling equation (\ref{eq:4}), one can
explicitly write the magnetic field as a function of $f$

\begin{equation}\label{bgen}
  \B= \frac{B_{pole}}{2} \left(\frac{r}{R_{NS}}\right)^{-p-2}
  \left[-f', \frac{pf}{\sin\theta},
    \sqrt{\frac{C\ p}{p+1}} \frac{f^{1+1/p}}{\sin\theta}\right]
\end{equation}
where a prime denotes derivation with respect to $\mu$.
Finally, taking into account equations (\ref{ppdep}) and
(\ref{fp}), the GSS equation becomes for the case at hand

\begin{equation}
  \label{ODE}
  (1-\mu^2)f'' + p\ (p+1)f + C f^{1+2/p}=0
\end{equation}
which is a second order ordinary differential equation for the angular part of
the flux function. Its solution, once a suitable set of boundary conditions has
been supplied (see \S\ \ref{bc}), completely specifies the external magnetic field.

Besides controlling the radial decay, the parameter $p$
also fixes the amount of shear of the field. In fact, recalling the definition of shear
angle (W95, TLK)
\begin{eqnarray}\label{twistangle}
  \Delta \phi_{NS} &=&
  \int_{\textrm{field line}} \frac{B_\phi}{(1-\mu^2) B_\theta}\ud\mu
  \nonumber \\
  &=& \left[\frac{C}{p\ (1+p)}\right]^{1/2}
  \int_{\textrm{field line}} \frac{f^{1/p}}{1-\mu^2} \ud\mu \, ,
\end{eqnarray}
it is immediate to see  that different
values of $p$ correspond to fields with different shear. Actually, as it will be
discussed later on, the effect of decreasing $p$ is to increase $B_\phi$ with respect to
the other components, and consequently to increase the shear.

As it is apparent from equation (\ref{bgen}), the limiting case $p \to 0$ results in a
purely radial field whose field lines are directed
either outwards or inwards (whence the name of split monopole).
The directions of the field lines divide the sphere (i.e. the star)
into several zones,
and we can imagine field lines of opposite directions connecting at radial infinity
(W95). Split monopole fields obtained for different multipolar orders
differ from one another in the number of zones into which the
sphere is split.

\begin{figure}
  \centering
  \includegraphics[width=0.22 \textwidth]{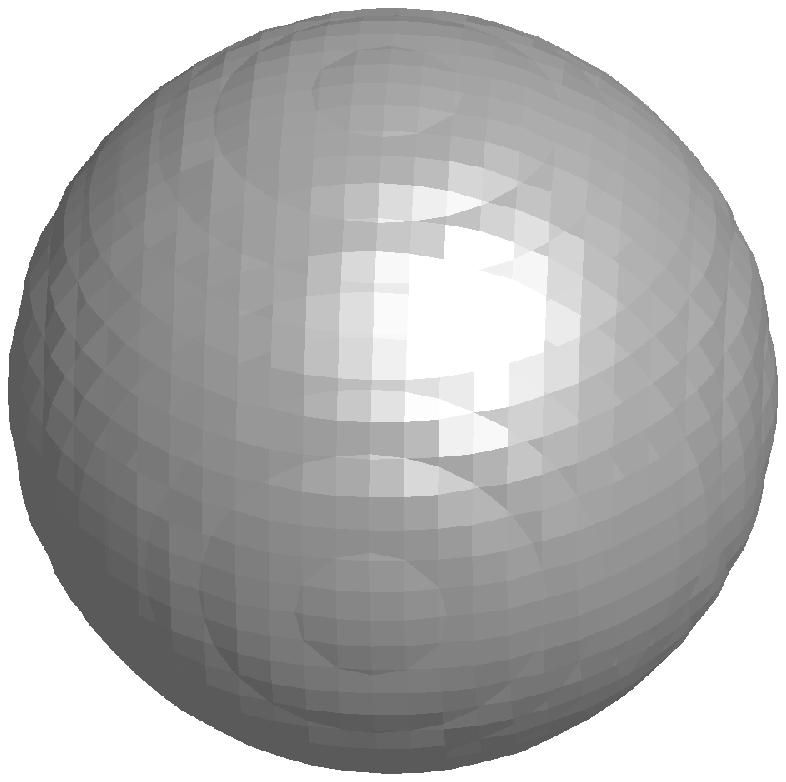}
  \includegraphics[width=0.22 \textwidth]{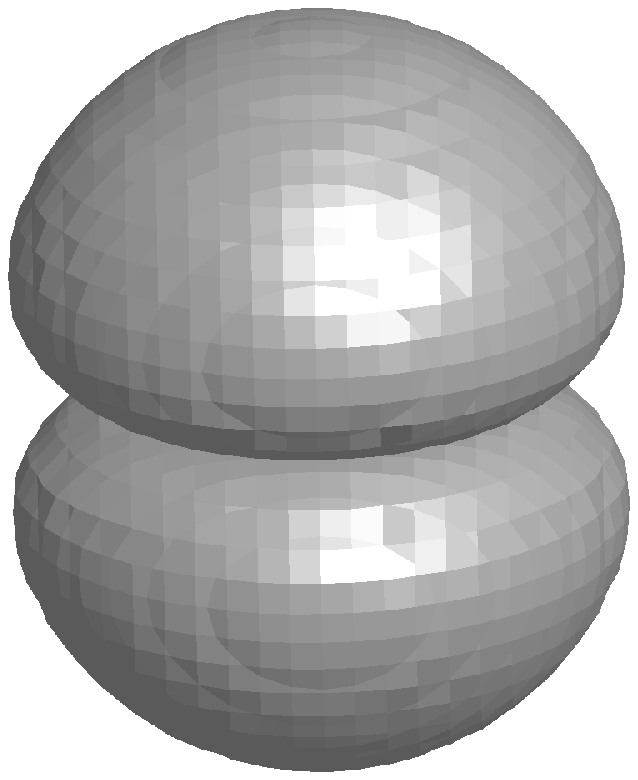}
  \includegraphics[width=0.22 \textwidth]{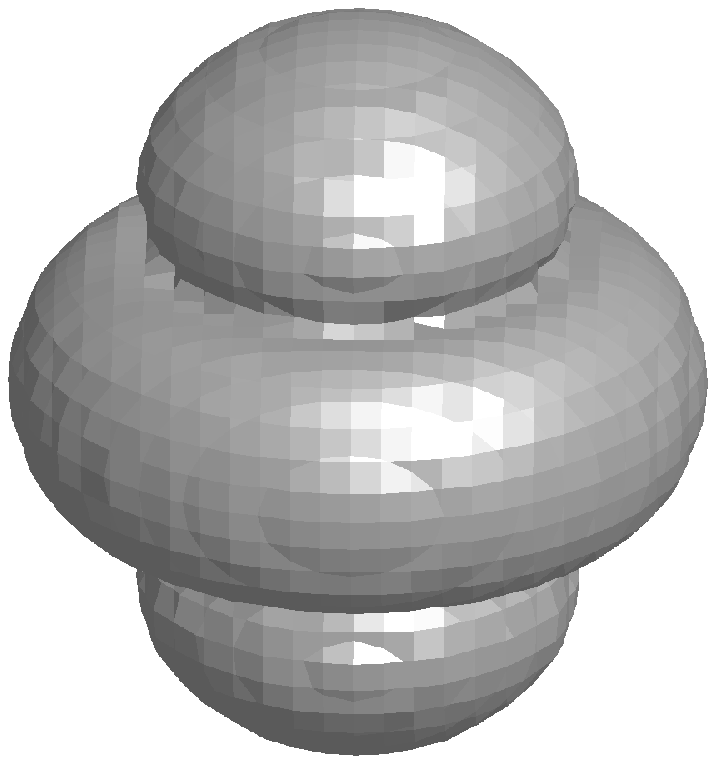}
  \includegraphics[width=0.22 \textwidth]{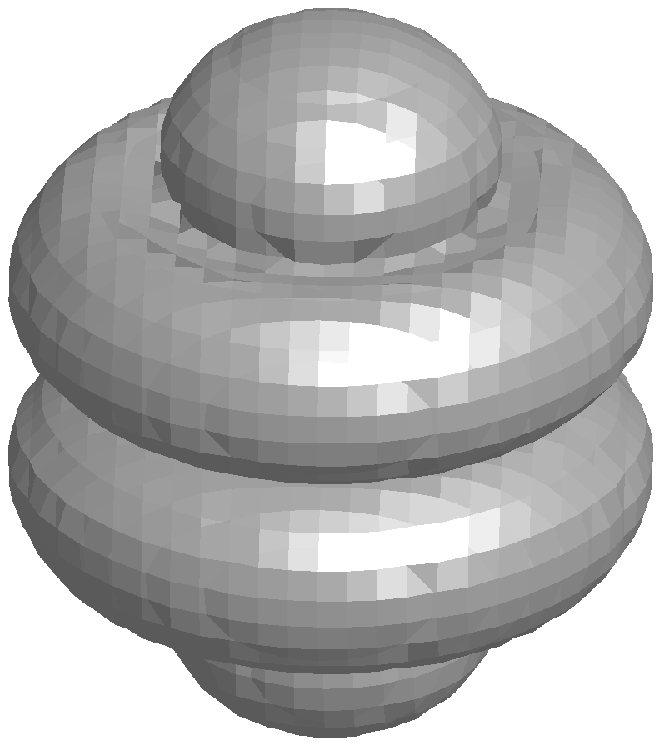}
  \caption{Zonal harmonics are used to
    visualize multipolar field topologies. From top left to bottom
right: multipoles with order from one to four.
    The positions of the degenerate poles are evident.}
\label{fig:magnetic-hemispheres}
\end{figure}

\subsection{Boundary conditions}\label{bc}

In order to solve equation (\ref{ODE}) we need to provide a suitable set of boundary conditions.
For a dipolar field there are just two poles and the star is divided into two hemispheres,
i.e. the portions of the surface $0\le\theta\le\pi/2$, $\pi/2\le\theta\le\pi$,
respectively, and $0\le\phi\le 2\pi$. Higher order multipoles have in addition degenerate poles,
which are loci of constant colatitude, and correspond to circles on the surface (see
Fig. \ref{fig:magnetic-hemispheres}).
For the quadrupole, for example, there are two poles on the magnetic axis and the equator is
a degenerate pole.
In the case of a generic multipolar field, the magnetic hemispheres of the dipole are replaced
by a number of regions, each limited by
two consecutive values of the colatitude, $\theta_i$, $\theta_{i+1}$, for which the field is purely radial
(the pole) and has vanishing radial component (the magnetic ``equator''), i.e. $B_\theta(\theta_i)=0$ and
$B_r(\theta_{i+1})=0$. We refer to these zones as to regions of unipolar $B_r$.

Because of the north-south symmetry of unsheared multipolar fields, which is assumed to hold
also for their twisted counterparts, the integration domain is restricted to $0\leq\mu\leq 1$. In analogy with
the ``composite magnetic fields'' of
\cite{low+lou}, integration is performed piecewise, going from one pole to the next one.
In each interval, the boundary conditions for equation (\ref{ODE}) are determined by
the requirements that: i) $\B$ is purely radial at each pole, and ii)
the intensity is not modified by the shear.
The latter condition can be enforced either by assigning the field strength at
one pole, or the magnetic flux out of each region of unipolar $B_r$; in both cases the value must
not change with the shear.
For multipoles of odd
order (dipole, octupole, \ldots) the first sub-domain starts at the geometrical equator, $\mu=0$, which is not a pole.
In these cases the boundary condition
at $\mu=0$ reflects the N-S symmetry of the field and translates into $f'(0)=0$ (TLK).

The magnetic field in the entire interval ($0\leq\mu\leq 1$) is then obtained
by assembling the solutions computed in the various sub-domains with the same value of $p$,
to ensure that the radial dependence of $\B$ is the same at all co-latitudes.
Since the GSS equation is a second order ODE, imposing three boundary conditions
implies that the parameter $C$ is an eigenvalue of the problem and depends on the radial index $p$, $C=C(p)$.
Each multipolar potential term satisfies the GSS equation with $C=0$ and is associated to
an integer radial index $p_0$; it is $p_0=1$ for the dipole, $p_0=2$ for the quadrupole and so on.
In this particular case, the equation is linear and  admits analytical solutions of the form (W95)

\begin{equation}\label{untwisted}
  f_{p_0}(\mu) \propto \sqrt{1-\mu^2}\ P^1_{p_0}(\mu),
\end{equation}
where $P^1_p(\mu)$  is the associated Legendre function of the
first kind \citep{abramowitz}.

Having established a set of boundary conditions, the GSS equation can
be solved for different values of $p$, building a sequence of models characterized by
a varying shear. As shown by \cite{low+lou}, such a sequence share the same topology,
i.e. the same number and position of poles. The only permitted values of the radial index are
$p \le p_0$.

\subsection{Dipolar fields}\label{sec:dipolar-fields}

The generating
function of a pure dipole is
\begin{equation}
  f_{p_0=1}= 1-\mu^2,
\end{equation}
thus
from equation (\ref{bgen}) it follows
that the condition at the pole $\mu=1$ is $f(1)=0$, to which the symmetry condition
$f'(0)=0$ must be added. The third condition is set either
specifying the field strength at one pole,
$f'(1)=1$ (as in TLK),
or requiring  a constant
flux, $f(0)=1$ (W95). The sequence of sheared dipoles has radial index $0\le p\le 1$.

\begin{figure}
\begin{center}
\includegraphics[width=0.44 \textwidth]{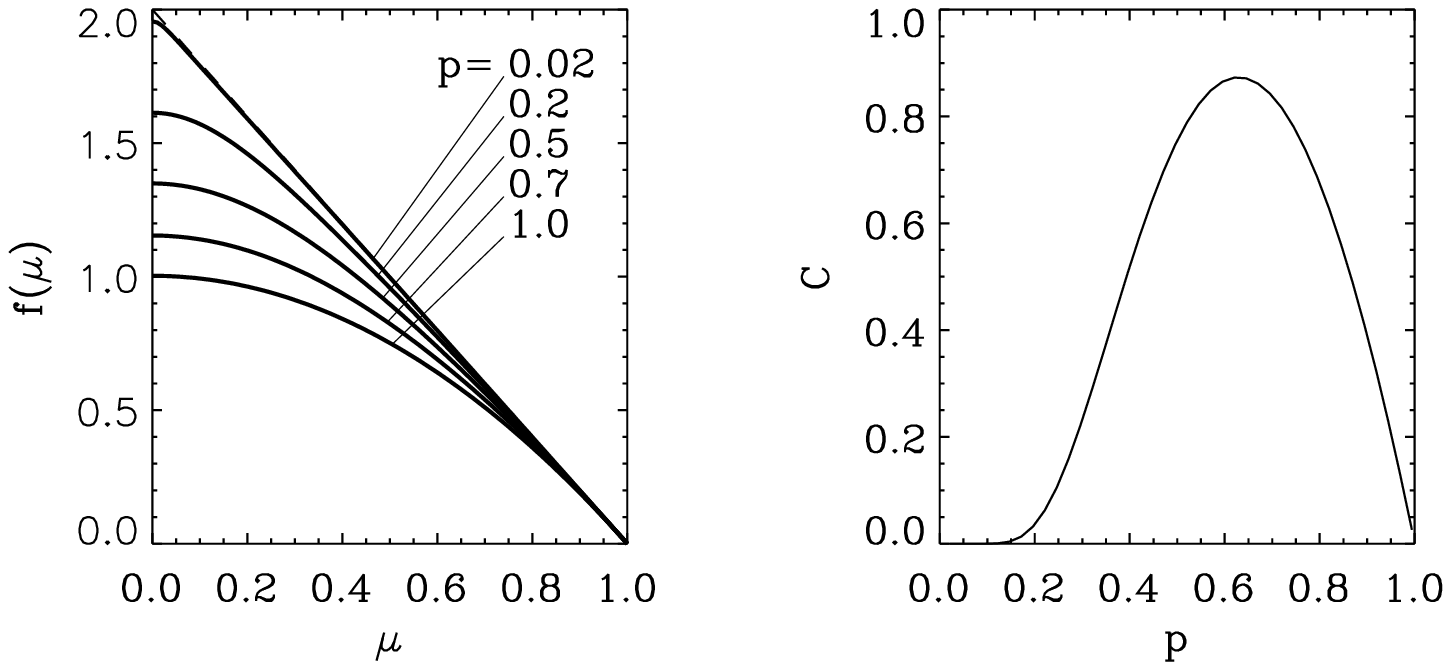}
$\phantom{.}$  
\includegraphics[width=0.44\textwidth]{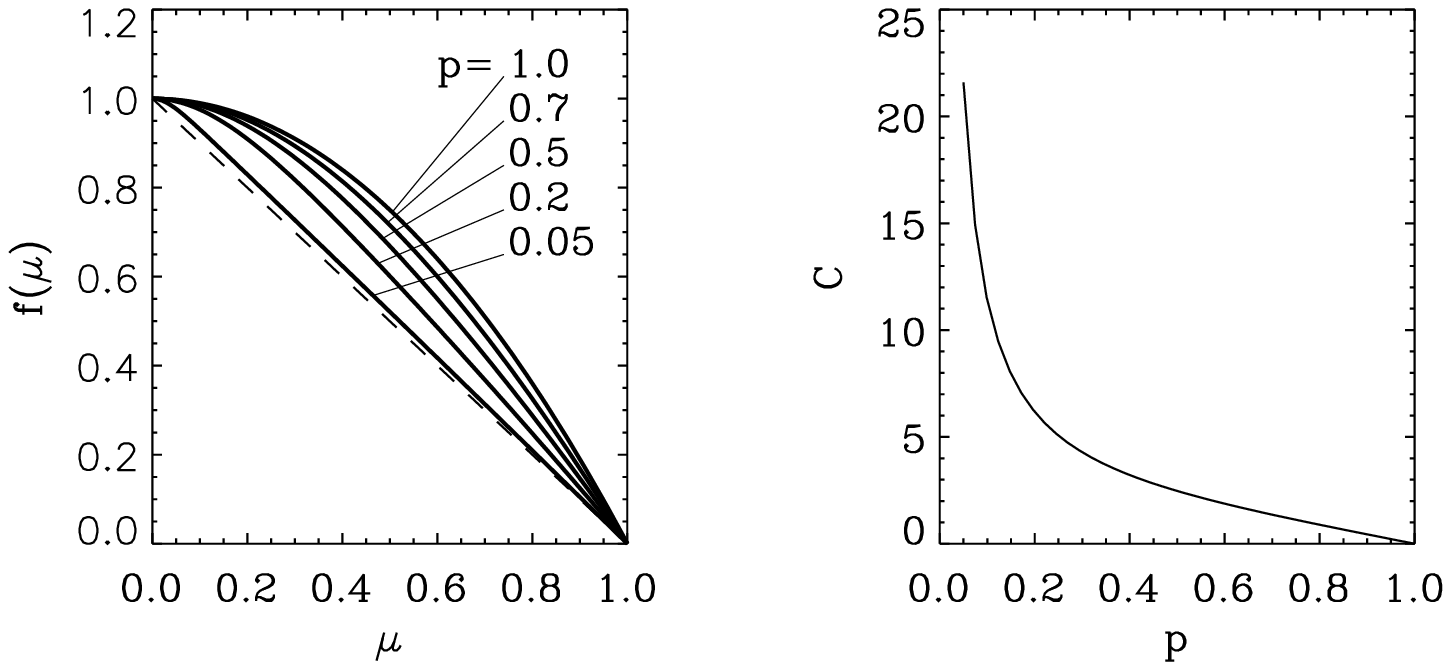}
\end{center}
\caption{Upper panels: the function $f(\mu)$ and the eigenvalue $C$ for different values of $p$
with the prescription of constant $B_{\textrm{pole}}$ (the dashed line in the left panel is the analytical
solution for $p \to 0$).
Lower panels: same, but for constant flux.}
\label{func_wolf}
\end{figure}

Sheared dipole fields have been discussed by W95 and TLK.
These investigations, however, used a different boundary condition (see above)
and we verified that the numerical solution of equation (\ref{ODE}) produces quite different
results for $f(\mu;\, p)$ and even more diverse eigenvalues $C(p)$ in the two cases (see
Fig.~\ref{func_wolf}). Still, one expects the magnetic field to be the same
in both cases, since the two boundary conditions are physically equivalent. Actually, a direct
comparison of the numerical solutions shows that $f_{TLK}/f_{W95}\simeq f'_{TLK}/f'_{W95}\simeq {\rm constant}$
for each $p$. Denoting by $\lambda(p)$ this constant ratio, one can write the expression of the field
in the two cases using equation (\ref{bgen})

\begin{eqnarray}
\nonumber
  \B_{W95} & \propto& \left[-f', \frac{p}{\sin\theta}
    f, \left(C_{W95}\ \frac{ p}{p+1}\right)^{1/2}
    \frac{f^{1+1/p}}{\sin\theta}\right] \\
  &&\\
\nonumber
  \B_{TLK} &\propto& \left[- \lambda f',  \lambda \frac{p}{\sin\theta}
    f, \left(C_{TLK}\ \frac{ p}{p+1}\right)^{1/2}
    \frac{ (\lambda\  f)^{1+1/p}}{\sin\theta}\right].
\end{eqnarray}
where the radial dependence has been omitted. The two fields have the same topology if and only if
\begin{equation}\label{eq:7}
  \lambda^{1+1/p}\ \sqrt{C_{TLK}}  = \lambda\  \sqrt{C_{W95}}.
\end{equation}
We checked that the previous condition is indeed satisfied by our numerical solutions with a relative accuracy $\sim 2$\%
for nearly all values of $p$~\footnote{The error becomes larger for $p\sim p_0$ because $\lambda~\sqrt{C_{W95}}\to~0$}.
The eingevalues and the ratio $\lambda$ for different values of $p$ are reported in table \ref{tab:lambda}.

\begin{table}
  \centering
  \begin{tabular}{c | c | c | c| c}
    $p$  & $C_{TLK}$& $C_{W95}$& $\lambda$\\
\hline
      0.97  & 0.13 & 0.11& 1.02 \\
      0.89  & 0.41 & 0.45 & 1.06 \\
      0.82  & 0.63 & 0.79 & 1.10 \\
      0.74  & 0.78 & 1.14 & 1.15 \\
      0.67  & 0.86 & 1.51 & 1.21 \\
      0.59  & 0.86 & 1.91 & 1.26 \\
      0.52  & 0.79 & 2.35 & 1.33 \\
      0.45  & 0.64 & 2.87 & 1.39 \\
      0.38  & 0.43 & 3.52 & 1.47 \\
      0.30  & 0.22 & 4.41 & 1.56 \\
      0.22  & 0.06 & 5.80 & 1.65 \\
      0.15  & $3.9 \times 10^{-3}$ & 8.52& 1.75 \\
     $0.07$ & $4.3 \times 10^{-7}$ & 17.48 & 1.88 \\
     $0.02$  & $1.55 \times 10^{-25}$ & 89.78 & 1.97 \\
\hline
\end{tabular}
\caption{The eigenvalues $C(p)$ for the two different sets of boundary conditions and the ratio
$\lambda =f_{TLK}/f_{W95}$ for $0< p<1$.}\label{tab:lambda}
   \end{table}

The magnetic field, then, has the same topology in both cases, and the two solutions
differ only for a multiplicative factor (which changes with $p$); i.e., it is
$\B_{{TLK}} = \lambda \B_{{W95}}$, with $1 \leq \lambda\leq 2$.
The great diversity in the eigenvalues, especially for $p\to0$
is in fact balanced by the different behavior of the function $f(\mu)$.
While for constant $B_{\textrm{pole}}$, it is $1 \le \max f \le 2$,
in the case of constant flux $\max f=1$ for any $p$. In the limit $p\to 0$ the
proportionality of the two solutions for $f$ can be recovered analytically. In fact, it
can be shown that
in the split-monopole limit the generating functions are
\begin{eqnarray*}
  & f(\mu)=1-|\mu|    & \textrm{for constant flux}\\
  & f(\mu)=2\ (1-|\mu|) & \textrm{for constant $B_{\textrm{pole}}$}
\end{eqnarray*}
which gives $\lambda(0)=2$, in agreement with the numerical result (see again table
\ref{tab:lambda}).

\subsection{Higher order multipoles}\label{sec:mult-solut}

\begin{figure}
  \centering
  \includegraphics[width=0.44 \textwidth]{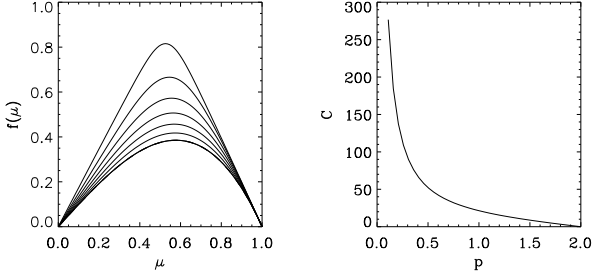}
  \caption{Generating functions and eigenvalues for the
    quadrupolar fields. Different curves correspond to
    different values of $p$ from $p=2$ (lower curve) to
    $p=0.2$ (upper curve). }
  \label{f_C_quadr}
\end{figure}

Since equation (\ref{ODE}) is the force-free condition
for a generic axisymmetric field (as given by equation [\ref{eq:4}]), axisymmetric globally-twisted multipoles
can be found solving again equation (\ref{ODE}), subject to the  boundary conditions
discussed in \S \ref{bc}.
For an untwisted quadrupolar field the generating function is
\begin{equation}
  f_{p_0=2}= \mu\ (1-\mu^2).
\end{equation}
Within our integration domain $0\le\mu\le 1$, the poles are located at $\mu=0$ (degenerate)
and $\mu=1$. The boundary conditions are
then
\begin{eqnarray}
  f(1)=0 & & \\
  f(0)=0 & & \\
  f'(1)=-2 & {\rm or}& f \left( \frac{1}{\sqrt{3}} \right) = \frac{2}{ 3 \sqrt{3}};
\end{eqnarray}
the latter two conditions enforce either constant field strength at the pole or
constant flux, and, as discussed in the previous section are
equivalent. However, on a numerical ground, we found that for higher order multipoles
the constant flux boundary condition is highly preferable and has been used in the calculations
presented here. The field in $-1 \le \mu \le 0$ is obtained by symmetry. The quadrupolar angular functions
$f(\mu)$ and eigenvalues $C(p)$ are shown in Fig. \ref{f_C_quadr} for different values of $p$.
\begin{figure}
  \centering
  \includegraphics[width=0.44 \textwidth]{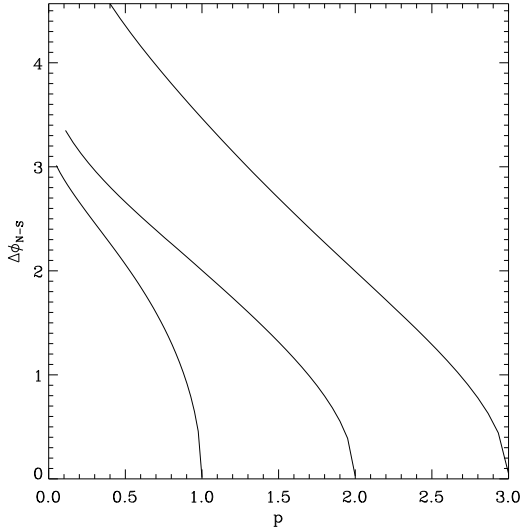}
  \caption{The shear angle $\Delta\phi_{NS}$
    as a function of $p$. The three curves correspond to
    dipolar  quadrupolar and octupolar fields (from left to
    right).}
  \label{deltaphi_ott}
\end{figure}
The shear angle as a function of $p$, equation (\ref{twistangle}), is shown in Fig.~\ref{deltaphi_ott}
(middle curve).

\begin{figure}
  \centering
  \includegraphics[width=0.44 \textwidth]{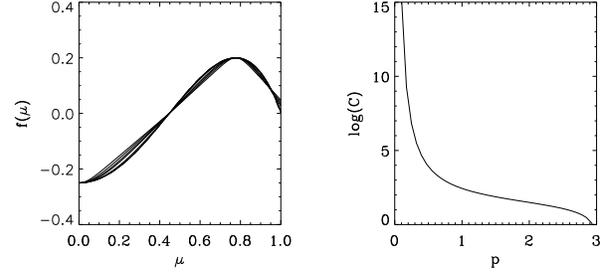}
  \caption{Same as Fig.~\ref{f_C_quadr} but for
    octupolar fields; here $p$ is in the range $[0.2, 3]$. }
  \label{f_C_ott}
\end{figure}

In the case of an untwisted octupole, instead, the
generating function is
\begin{equation}
  f_{p_0=3} = \frac{1}{4} (1-\mu^2) (5\mu^2 -1)
\end{equation}
and the poles are located at $\mu=1, 1/\sqrt{5}$ (again restricting
to the range $0\le\mu \le 1$).
In order to compute the sheared field, equation (\ref{ODE}) needs to
be solved in two separate intervals,
$0\le\mu\le 1/\sqrt{5}$ and $1/\sqrt{5}\le\mu\le 1$. Taking into account
that $\mu=0$ is not a pole in the present case, the boundary conditions are
\begin{eqnarray}
  f(1/\sqrt{5})=0 & &\\
  f'(0)=0 & &\\
   f'(1/\sqrt{5})=\frac{2}{\sqrt{5}} & {\rm or}& f(0)=- \frac{1}{4}
\end{eqnarray}
in the range $0 \le \mu \le 1/\sqrt{5}$ and

\begin{eqnarray}
  f(1)=0 & & \\
  f(1/\sqrt{5})=0 & &\\
  f'(1)=-2 & {\rm or}& f \left( \sqrt{\frac{3}{5}} \right) = \frac{1}{5}
\end{eqnarray}
in the range $1/\sqrt{5} \le \mu \le 1$. The numerical solutions are shown in
Fig. \ref{f_C_ott} and the dependence of the shear angle on $p$ can be read
from Fig.~\ref{deltaphi_ott} (rightmost curve).

Despite the numerical solution of equation (\ref{ODE}) poses no problems, having analytical expressions for the twisted
field components may prove handy in some applications. In Appendix \ref{sec:app} we derive
approximated analytical expressions for the generating
function $f(\mu)$ for dipolar and quadrupolar configurations valid for small shear ($p_0-p\ll 1$), compare them with the
exact numerical solutions and discuss their ranges of validity.

\section{Spectra and lightcurves}\label{sec:spectra}

In this section we use the numerical code described in
\citet[NTZ in the following]{ntz1} to explore  the effects of different
sheared multipolar fields on the emergent spectra and lightcurves of magnetars. 
 Although the external field of a magnetar
is unlikely to comprise a single higher order multipole, investigating
spectral formation when the field is one (twisted) multipole offers the
opportunity to explore the effects of magnetospheric currents localized on
spatial scales smaller than that implied by the (twisted) dipole. An example 
is that of a sheared, localized component which appears as a consequence
of some form of activity and adds up to a global, (quasi)potential dipolar
field. The currents responsible for the resonant scatterings are provided
only by the sheared field. The case in which the localized component is modeled in terms of the 
polar ``lobe(s)'' of an octupole, is discussed in \S\ref{sec:pulsed}.
We base our model on the scenario envisaged by TLK \cite[see also][for more detailed calculations]{lg06,fern,ntz1},
according to which thermal photons originating at the star surface undergo repeated
scatterings with the charge carriers (electrons, ions and possibly pairs;
see also \citealt{belotho07}) flowing along the field lines. These investigations were based on non-relativistic
computations, and therefore necessarily restricted to the low-energy ($\la
10$ keV) emission. Furthermore, they were based on the
dipolar, globally-twisted magnetosphere of TLK \footnote{The simplified, analytical
model of \cite{lg06} did not assume a precise topology for the magnetic
field.}.

\subsection{Globally twisted multipoles}\label{sec:spectra-glob}

\begin{figure}
  \centering
  \includegraphics[width=0.44 \textwidth]{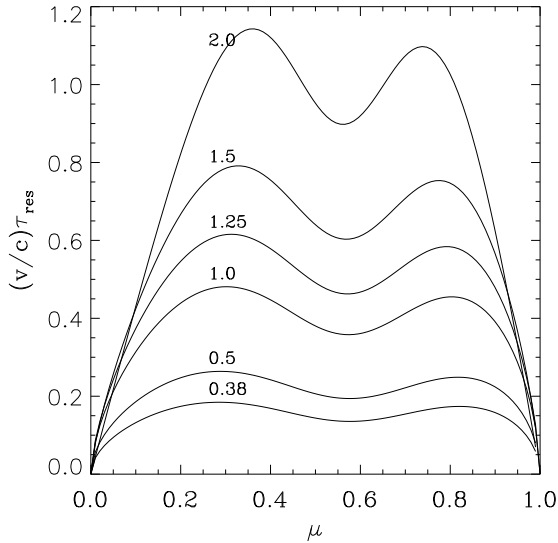}
  \caption{Optical depth to resonant cyclotron scattering at the resonant radius
    as a function of the colatitude for quadrupolar configurations;
    each curve is labelled by the value of the shear angle $\Delta\phi_{NS}$.}\label{fig:tauQuadr}
\end{figure}

\begin{figure}
  \centering
  \includegraphics[width=0.44 \textwidth]{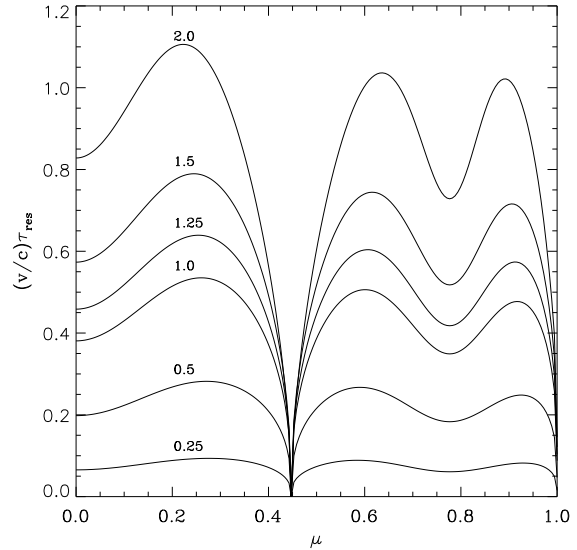}
  \caption{Same as in Fig. \ref{fig:tauQuadr} but for an octupolar field.}\label{fig:tauOtt}
\end{figure}

\begin{figure}
  \centering
  \includegraphics[width=0.44 \textwidth]{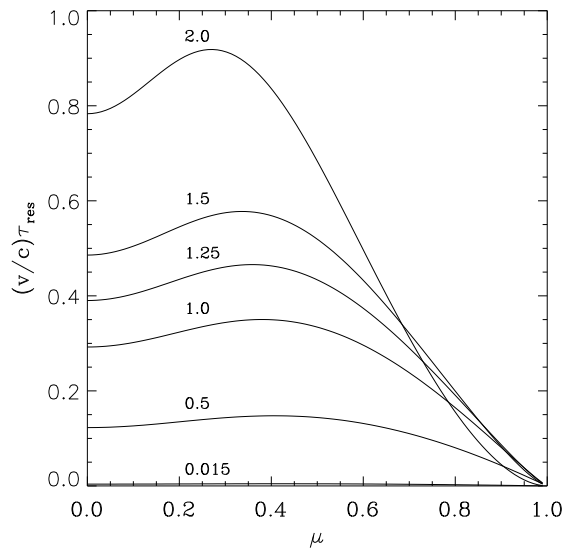}
  \caption{Same as in Fig. \ref{fig:tauQuadr} but for a dipolar field.}\label{fig:tauDip}
\end{figure}

In order to gain some insight on the properties in the emitted spectra when the
magnetosphere is threaded by twisted, higher order multipoles, we  plot in
Figs.~\ref{fig:tauQuadr} and \ref{fig:tauOtt} the
optical depth $\tau_{res}$ to resonant scattering corresponding to
quadrupolar and octupolar fields. This is given by
\begin{equation}\label{depth}
  \left(\frac{v}{c}\right)\tau_{res} \sim \frac{\pi}{4} (1+\cos^2\theta_{kB})
  \left[ \frac{C(1+p)}{p}\right]^{1/2}\frac{f^{1/p}}{2+p}
\end{equation}
where $v$ is the charge velocity and $\theta_{kB}$ is the angle between the primary photon (assumed
to move radially) and the magnetic field at the scattering
radius (TLK). Since $v\sim c$, $(v/c)\tau_{res}\sim \tau_{res}$ and  we make no further distinction between
these two quantities. By comparing these curves to those in  Fig. \ref{fig:tauDip}  we can clearly see
that the  optical depth for higher order multipoles is sensibly
different from that of a
dipolar configuration, implying that the overall spectral properties, and
in particular those of the pulse phase emission,  may
be significantly affected.
 We remark that the typical radius at which resonant scattering occurs is
different for the different multipoles. Assuming that the surface strength
is comparable, scattering in higher order multipolar fields takes place closer
to the star with respect to the dipole because the field decays faster with
radius.

To further investigate this, we calculated a number of synthetic spectra and
lightcurves by using the non-relativistic Monte Carlo code by NTZ.
\begin{figure}
  \centering
  \includegraphics[width=0.44 \textwidth]{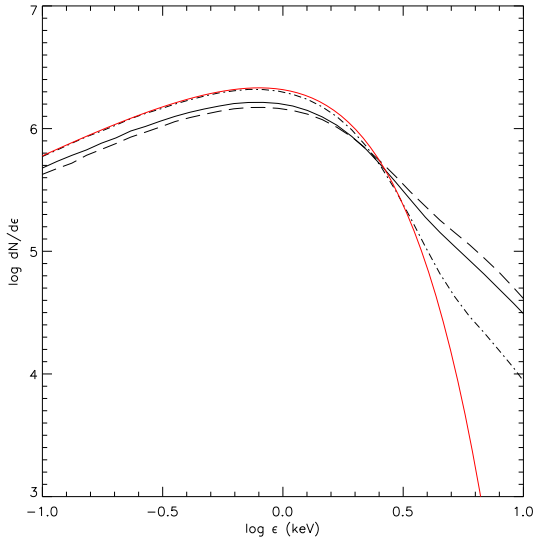}
  \caption{Monte Carlo spectra for globally twisted dipolar
    (solid line), quadrupolar (dashed line), octupolar
    (dash-dotted line) force free magnetospheres; the shear
    angle is the same in all three cases
    ($\Delta\phi_{NS}$=1.2 rad). The solid red line  is the seed
    blackbody spectrum. The total number of photons is $\sim 4\times 10^6$
    in all the simulations.}
  \label{spettri_tot1}
\end{figure}
A comparison among spectra produced by a globally twisted dipolar,
quadrupolar and octupolar magnetosphere is shown in Fig.~\ref{spettri_tot1}.
Spectra have been computed for the same values of model
parameters (blackbody temperature $kT_\gamma=0.5$ keV,
temperature and bulk velocity of the magnetospheric
electrons $T_{el}=30$ keV,
$\beta_{bulk}=0.5$, polar magnetic field strength
$B_{pole}=10^{14}$ G), and for the same shear angle
($\Delta\phi_{NS}=1.2$ rad, which corresponds to
$p=0.8,\, 1.6,\, 2.5$ for the dipole, the quadrupole and
the octupole, respectively)\footnote{ Although these values of the model 
parameters can be regarded as typical (see NTZ), the present
choice has only illustrative purposes. Other combinations of
the parameters are equally possible and, provided that electrons
remain mildly relativistic, will give similar results.}.
 Spectra have been computed by collecting photons over the
entire observer's sky (i.e. by angle-averaging over all viewing
directions). The most prominent feature in Fig.~\ref{spettri_tot1} is the higher
comptonization degree induced by the quadrupolar (and dipolar) field with
respect to the octupolar one. This can be understood in terms
of the different spatial distribution of the scattering
particles (see Figs.~\ref{fig:tauQuadr}, \ref{fig:tauOtt},  \ref{fig:tauDip}) and of the
different efficiency of scatterings, the latter depending on the (average) angle between
the photon direction and that of the flowing currents. Upscattering is more
efficient in regions where the currents move towards the star, i.e. close to the magnetic
south pole(s), because collisions tend to occur more head-on (see also NTZ). For a dipolar field
the more favourable situation (i.e. large optical depth and currents flowing towards
the star) arises for $\mu\sim -0.3$ (compare with the spectra
at different viewing angles in Fig. 1 of NTZ; the one at
$\Theta_S=116^\circ$ is the more comptonized).
For the quadrupole returning currents are localized around $\mu\sim 0$ (the geographical equator which
is a degenerate south pole) and there are two regions with large optical depth in $0\leq\mu\leq 1$ (see
fig.~\ref{fig:tauQuadr}). The one at $\mu\sim
0.3$ is closer to the south pole ($\mu=0$) than in the case of the dipole, for which
it occurs at $\mu\sim -0.3$ while the south pole is at $\mu=-1$. The reason for which octupolar twisted
fields produce less efficient upscattering is that, despite there are two maxima of the optical
depth  located quite close to the  south pole (at $\mu\sim 0.2,\, 0.6$, the south pole is at $\mu=1/\sqrt{5}\sim 0.45$),
the relatively large curvature of the field lines makes the angular extent of the region where currents are inflowing
narrow. As a consequence most photons scatter with electrons moving at large angles and this results in steeper spectra.

\subsection{A simple localized twist model}\label{sec:spectra-locl}

There is now observational evidence that, at least in some cases, the
magnetospheric twist in both SGRs and AXPs could
be localized in restricted regions of the magnetosphere. For
instance,  \cite{woods2} found a certain degree of hysteresis in the
long-term evolution of SGR 1806-20 prior the emission of the giant flare in December 2004, with
a non trivial correlation between spectral and  timing properties that may be
interpreted if only a small bundle of magnetic field lines is affected by
the shear. A further case is provided by the spectral evolution
of the transient AXP XTE J1810-197 \citep{perna, bern08} which seems to
require a twist concentrated towards the magnetic axis, giving rise to a polar
hot region, possibly with a meridional temperature gradient.

Motivated by this, we investigated the
possibility to model a localized twist, by constructing a solution in
which the shear changes with the magnetic colatitude of the field line
foot point. Although the configurations presented in the previous sections are globally twisted, the
octupolar solution, which requires piecewise
integration in two domains  (see \S\ref{sec:mult-solut}), can be used to construct a
field with a non-vanishing twist at low magnetic co-latitudes. This is done by superimposing a
sheared octupole for $0\le\mu\le 1/\sqrt{5}$ to a potential one in $1/\sqrt{5}\le\mu \le 1$
and is
equivalent to solve equation (\ref{ODE}) in the entire domain with $p(\mu)=p^*H(\mu-1/\sqrt{5})$, where $H$ is
the Heaviside step function and $p^*$ a given constant. We note that, although not physically self-consistent,
such a field is force-free (see the discussion in \S\ref{sec:conclusions}).

\begin{figure}
  \centering
  \includegraphics[width= 0.44 \textwidth]{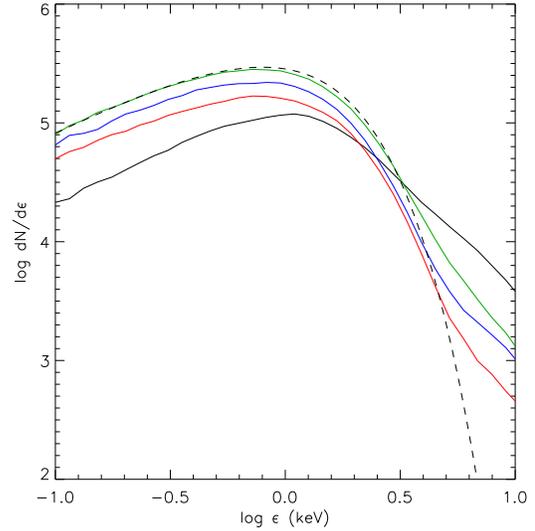}
  \caption{Spectra obtained with the Monte Carlo code for a
    locally twisted ($\Delta\phi_{NS}=1.5$~rad, $p=2.3$) octupolar force-free
    magnetosphere.  The shear is
    equally distributed on both polar lobes, while in
    the equatorial zone the magnetosphere is potential. Different curves correspond to different
    viewing angles (blue $\Theta_S=0^\circ$, red  $\Theta_S=60^\circ$, green
    $\Theta_S = 104.5^\circ$, black $\Theta_S=180^\circ$; the dashed line is the seed
    blackbody spectrum.
    }
  \label{spettro_2lobi}
\end{figure}

\begin{figure}
  \centering
  \includegraphics[width= 0.44 \textwidth]{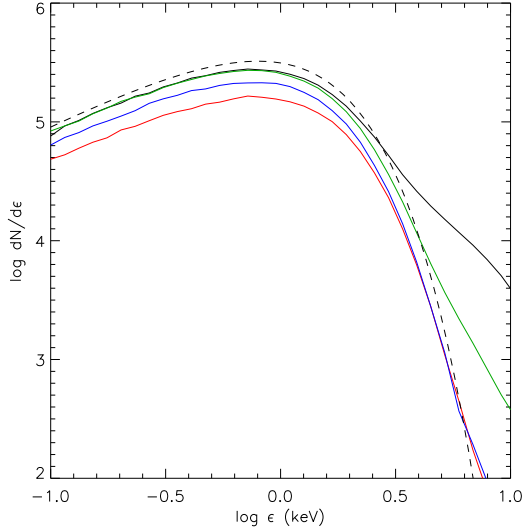}
  \caption{Same as in Fig.~\ref{spettro_2lobi} but for the shear localized only at
    upper north  pole.}
  \label{spettro_1lobo}
\end{figure}

The Monte Carlo spectra for different viewing angles are shown in
Fig.~\ref{spettro_2lobi} in the case in which the field has equatorial symmetry, i.e. the shear
$\Delta\phi_{NS} = 1.5$~rad ($p=2.3$) is  applied on both polar lobes
while  the equatorial zone is permeated by a
potential octupole. The spectra for the more interesting case in which the same shear is confined
only around one pole are plotted in Fig.~\ref{spettro_1lobo}.

\subsection{Timing and spectral properties of magnetars high-energy emission} \label{sec:pulsed}

Spectral models which account for different magnetospheric configurations hold the
potential to reproduce not only the gross features of
the observed spectra but also the subtler properties which are revealed by the combination of very
high-quality spectral and timing data. While a complete application is beyond the scope of this paper,
here we consider our results in the context of the spectacular phase-dependence which has
been recently discovered  in the the hard X-ray tails of the two AXPs 1RXS
J1708-4009 and 4U 0142+61 \citep{dhartog1, dhartog2}. These deep {\em INTEGRAL} observations have shown that,
in both these sources, there are several different pulse components (at least three) with genuinely different spectra.
The hard X-ray spectrum gradually changes with phase from a soft to an hard power law, the latter being
significantly detected over a phase interval covering $\sim 1/3$, or more, of the period.

\begin{figure}
  \centering
  \includegraphics[width= 0.44 \textwidth]{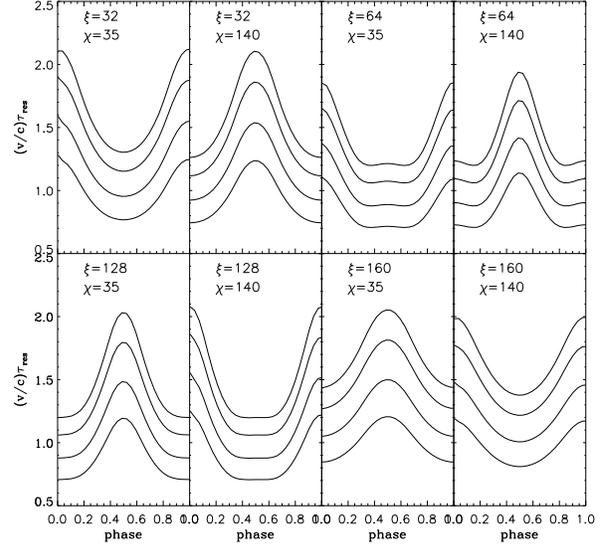}
  \caption{The depth $(v/c)\tau_{res}$ versus rotational phase for an
    octupolar field with twist located on the two polar caps. Each panel refer to different
  values of the geometrical angles $\chi$ and $\xi$ (see text for details). The curves in each panel
  are for $\Delta\phi_{N-S}= 1.22, 1.50, 1.78, 1.96$ from bottom to top.}
  \label{tau_phase}
\end{figure}

In order to see how these features can, at least qualitatively, be explained by our models, let us consider the
locally twisted configurations discussed in \S~\ref{sec:spectra-locl} which, among those presented so far,
provide the most significant variations of the magnetic topology with the colatitude. Let us introduce two
angles, $\chi$ and $\xi$, which give, respectively, the inclination of the LOS and of the magnetic axis with
respect to the star spin axis. This allows us to take into account for the star
rotation and hence derive pulse profiles and phase-resolved spectra. Because of the lack of north-south symmetry,
it is $0\leq\chi\leq\pi$,
while $\xi$ spans the interval $[0,\,\pi/2]$ (see NTZ for further details). For each viewing geometry, and for
different values of the shear, we can now compute the optical depth (equation~[\ref{depth}])
as a function the rotational phase $\gamma\ (0 \leq \gamma \leq 2 \pi)$. A few
examples are shown in Fig.~\ref{tau_phase}
for the case in which the  twist is localized on two polar caps.

\begin{figure}
  \centering
  \includegraphics[width= 0.44 \textwidth]{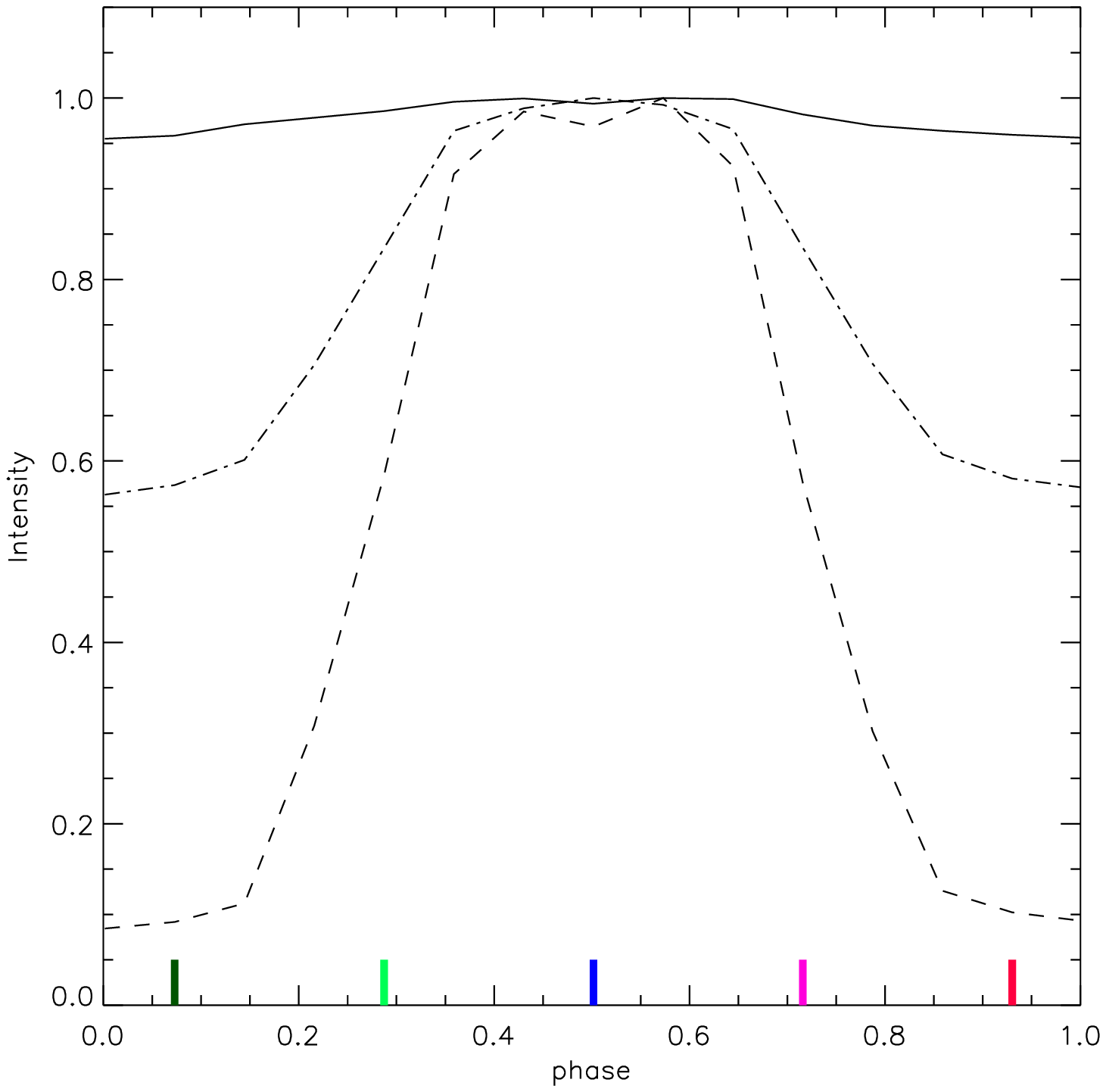}
  \includegraphics[width= 0.44 \textwidth]{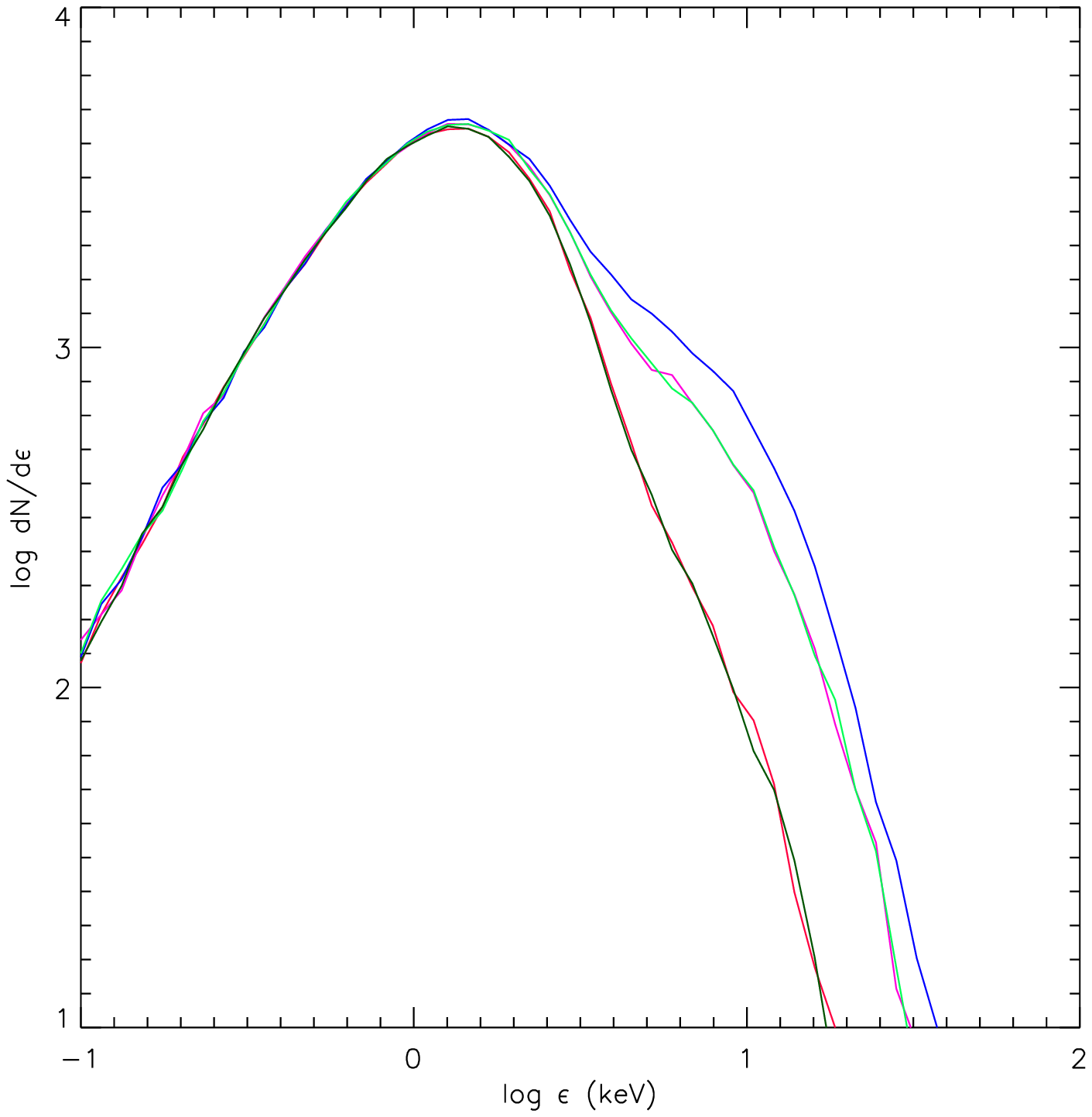}
   \caption{Synthetic lightcurves (upper panel) and
    phase-resolved spectra (lower panel) for a twisted magnetosphere
    with shear applied only to one polar lobe of an octupolar field
    (see text for details).
    In the upper panel the solid, dash-dotted and dashed lines refer to
    the pulse profiles in the 0.5--2, 2--10 and  10--100~keV bands respectively.
    In the lower panel curves with different colors give the spectra at various phases; the color code
    can be read  at the bottom of the upper panel.}
  \label{fig:spectra_phase}
\end{figure}

In connection with the spectral evolution with phase observed in 1RXS
J1708-4009 and 4U 0142+61, the  more favourable cases are those in which it is
$(v/c)\tau_{res} > 1$ for roughly one third of the period. This is because
resonant scattering over a population of (relativistic) electrons is then expected to be most efficient
in producing a hard tail over the right phase interval, while the decrease of the depth at other phases
results in a softening of the spectrum. A complete exploration of the parameter space 
aimed at searching for all configurations for which the previous condition 
is met is beyond the purposes of this paper. Just for illustrative 
purpose, let us consider one of those, i.e. an octupolar field with shear 
$\Delta\phi_{N-S}= 1.5$. By assuming this value and taking $\xi = 32^\circ $
$\chi = 140^\circ $, we then computed phase resolved
spectra and energy dependent lightcurves by using our Monte Carlo code. Since we are dealing with
photon energies $\approx 100$ keV, at which electron recoil and relativistic effects may become
important, we performed the runs using a relativistic version of the code
\cite[][Nobili, Turolla \& Zane, in preparation]{ntz2}. Results are
reported in Figs.~\ref{fig:spectra_phase} and \ref{fig:spectra_phase2lobi}. Again in the spirit of
probing the potentialities of our model, rather than a presenting a detailed fit, the
 parameter values in the Monte Carlo runs are the same as those adopted in
\S\ref{sec:spectra-locl}, $B_{\textrm pole}=10^{14}$ G, $\beta_{{bulk}}=0.5$, $T_{el}=30$~keV and $T_\gamma=0.5$~keV.
 We remark again that these values are not preferential, and that the results presented below do not critically
depend on the model parameters.

As it can be seen from Fig.~\ref{fig:spectra_phase}, for a magnetic configuration with the shear concentrated
in a single lobe, resonant comptonization gives rise to a hard tail which is quite pronounced at the
peak of the pulse while it is depressed by almost an order of magnitude at pulse phases close to the
minimum of the hard X-ray lightcurve. This is similar to what observed in AXP 1RXS J1708-4009 and
AXP 4U0142+61 \citep{dhartog1, dhartog2}. This model also predicts a
considerable variation of the pulsed fraction with energy, ranging from a few  percent below 2~keV, to a few
tens of percent from 2 to 10 keV and up to $90\%$ in the harder part
of the spectrum.
The comparison between modelled and observed phase-resolved spectra
is beyond the scope of the present investigation.

\begin{figure}
  \centering
  \includegraphics[width= 0.44 \textwidth]{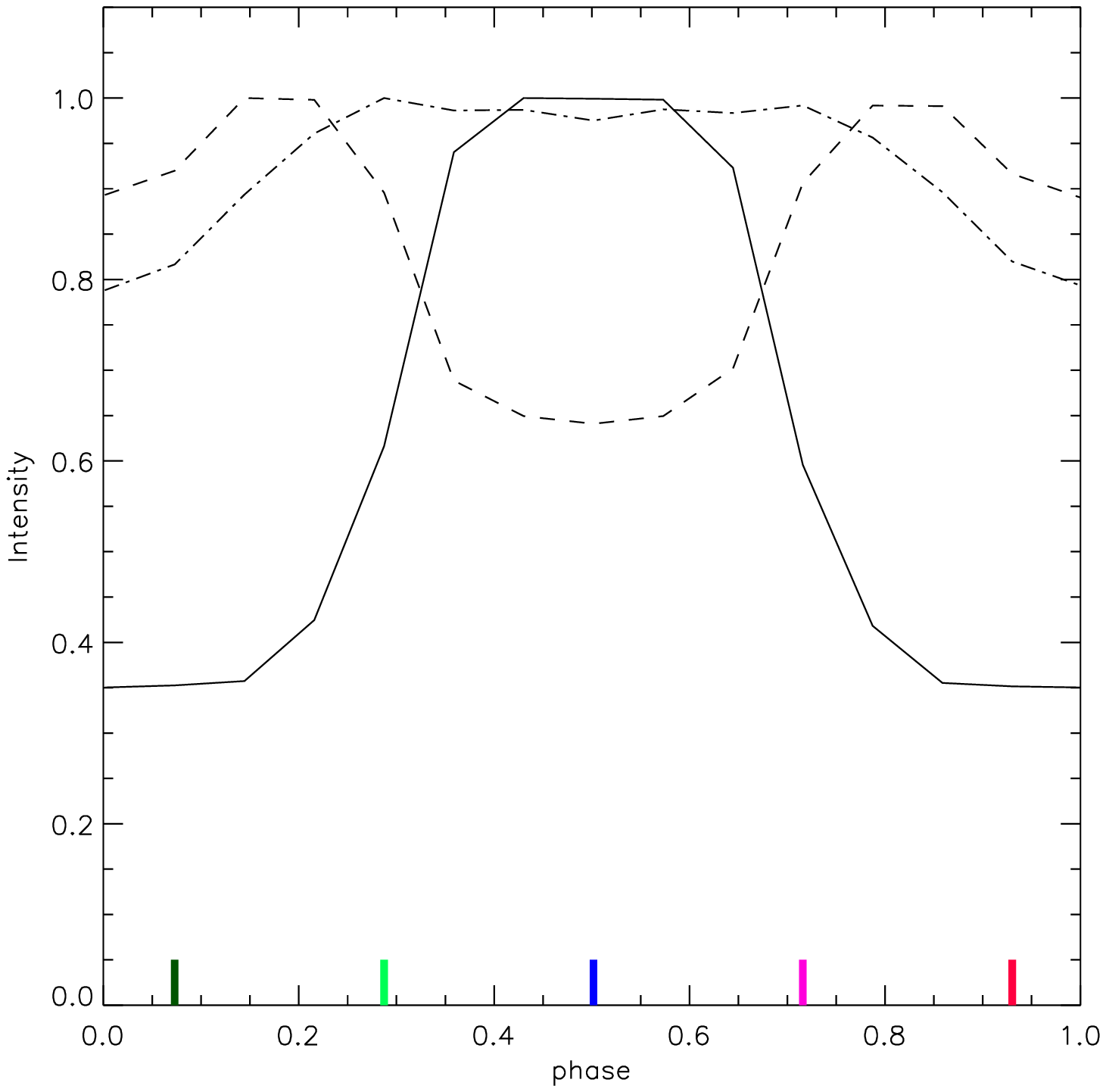}
  \includegraphics[width= 0.44 \textwidth]{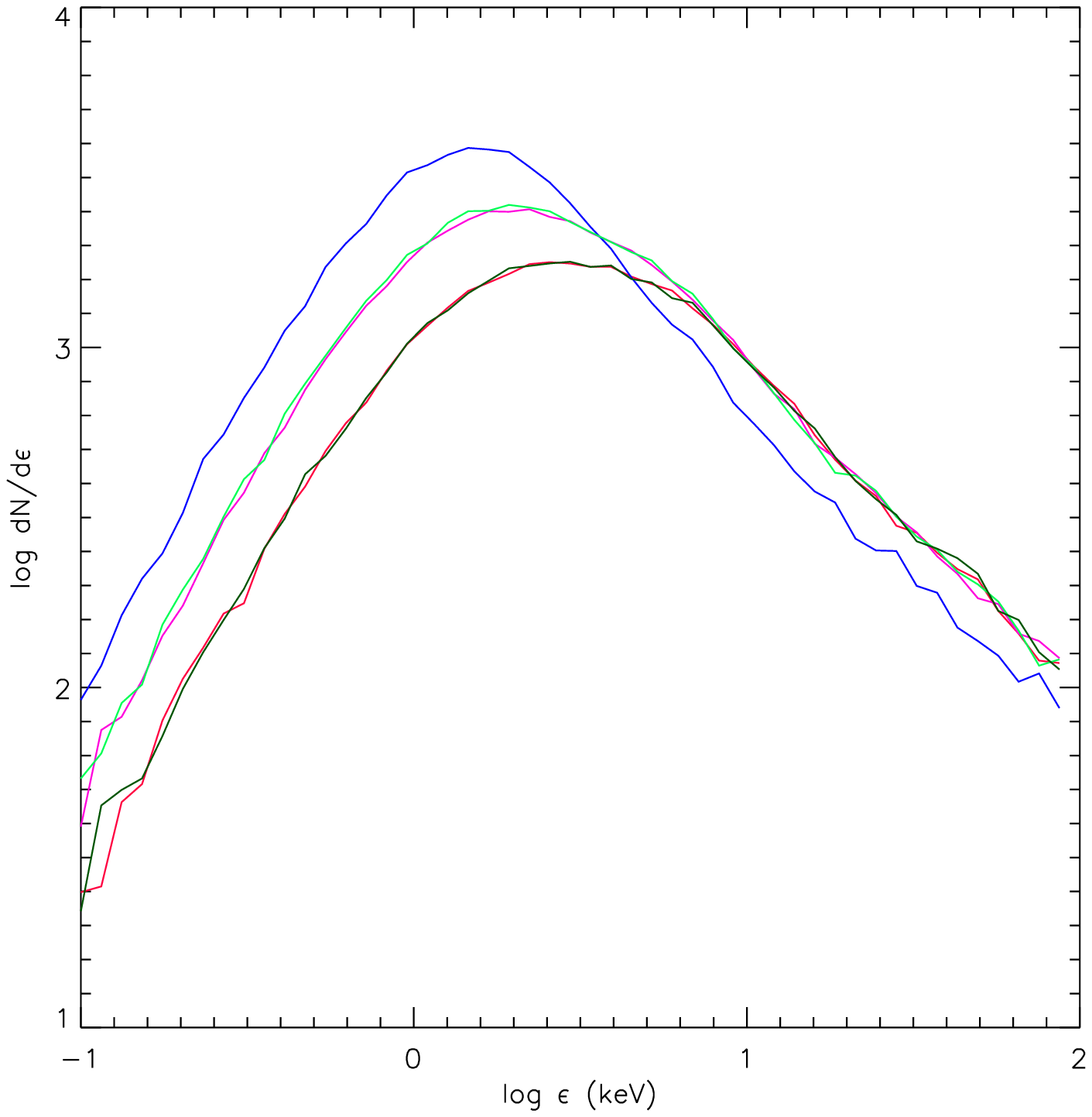}
  \caption{Same as Fig.~\ref{fig:spectra_phase} for
    a globally twisted dipolar field.}
  \label{fig:spectra_phase_dip}
\end{figure}

\begin{figure}
  \centering
  \includegraphics[width= 0.44 \textwidth]{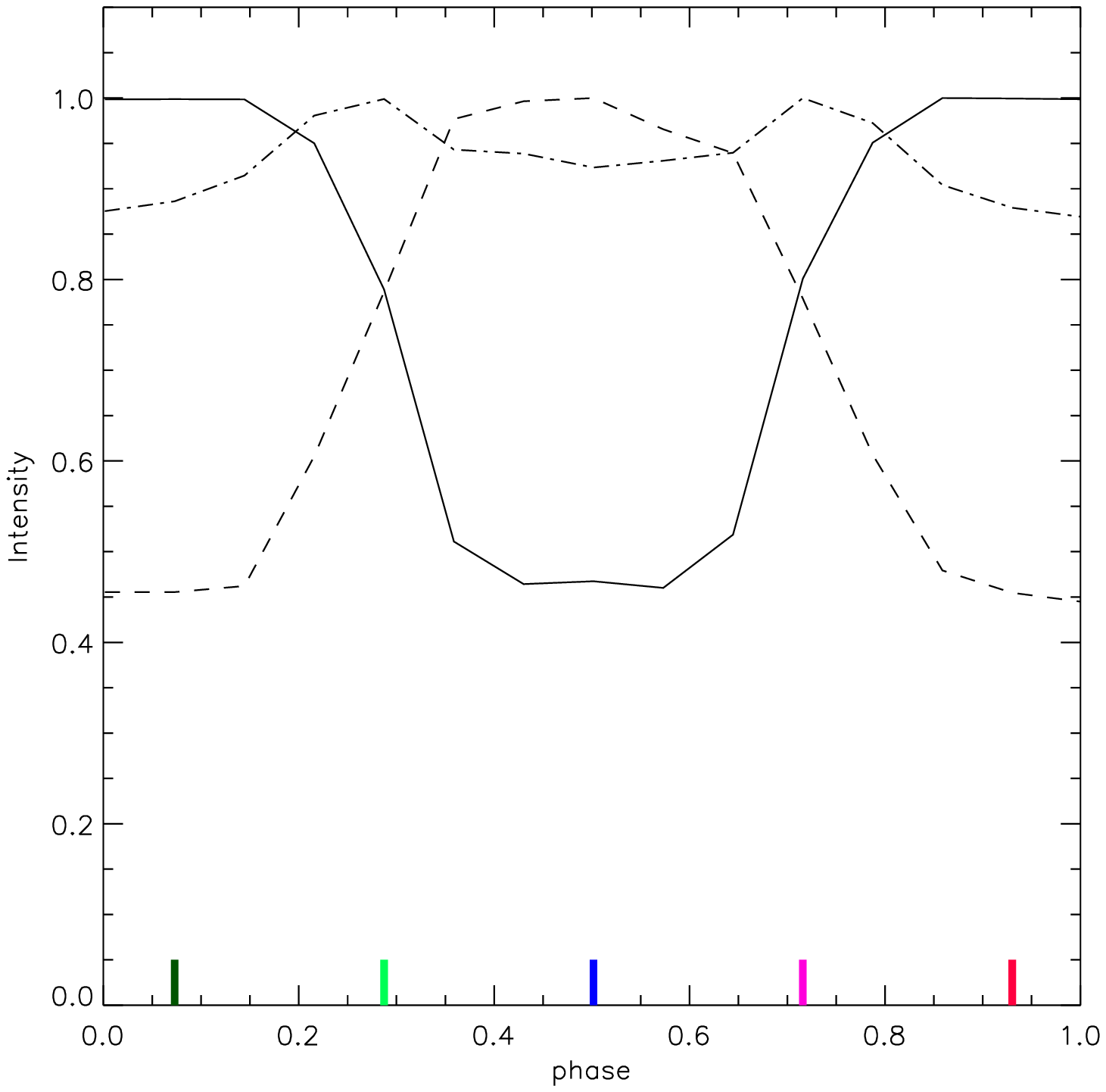}
  \includegraphics[width= 0.44 \textwidth]{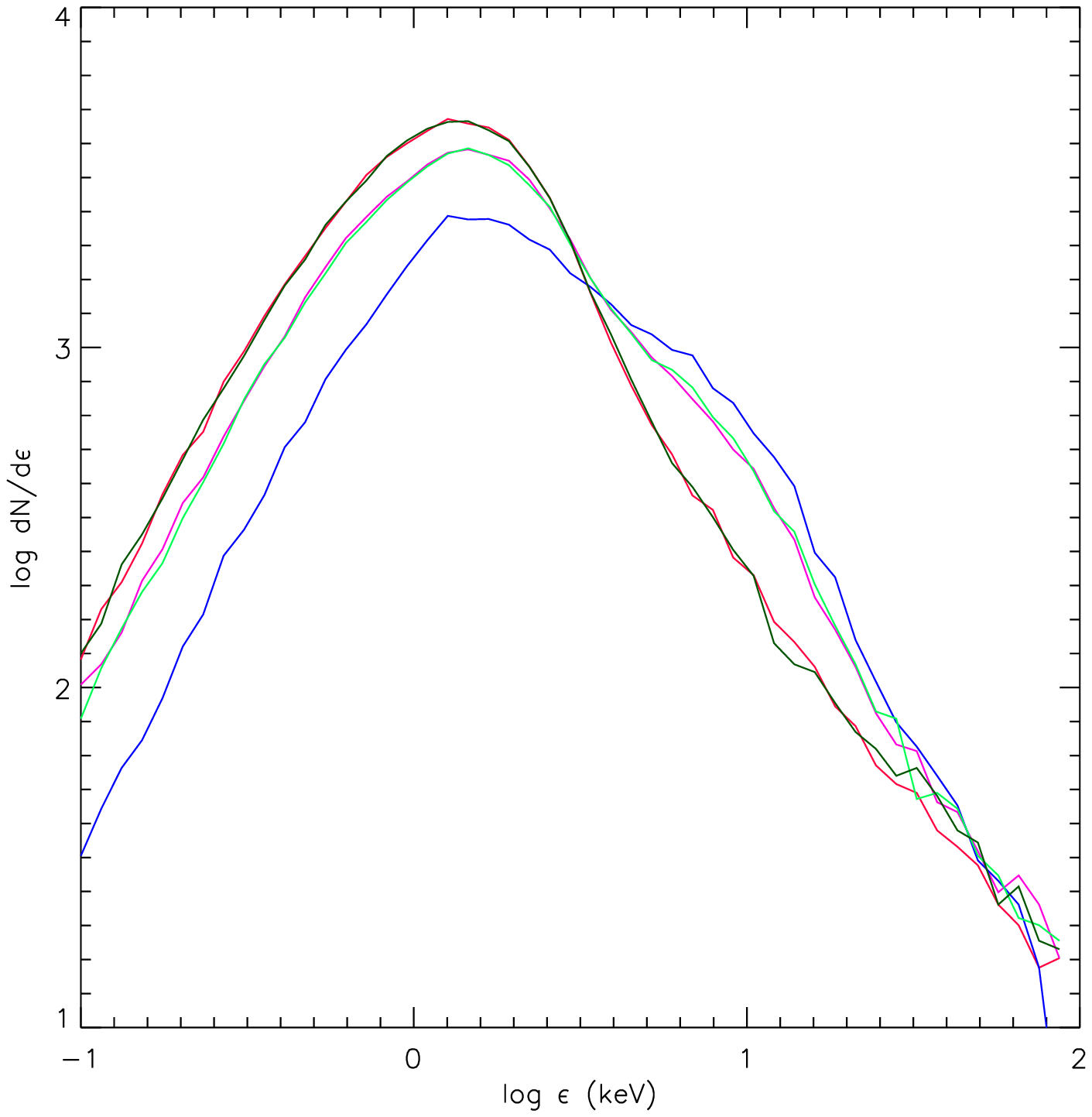}
  \caption{Same as Fig.~\ref{fig:spectra_phase} for the case in which the shear
  is applied to both polar regions.}
  \label{fig:spectra_phase2lobi}
\end{figure}

For comparison we show in Fig.~\ref{fig:spectra_phase_dip} the same results computed by assuming a
globally twisted dipolar magnetosphere, with the same shear and taking the same parameters for the viewing
geometry and the Monte Carlo run. It is clear that the pulse resolved
spectra and lightcurves obtained in this case are completely
different and can not reproduce those observed from the two AXPs. In this case, the hard part of the spectrum show
very little variation with the rotational phase and the lightcurve dependence on the energy is opposite to that
shown in Fig.~\ref{fig:spectra_phase}, with a larger pulsed fraction expected in the soft band. Spectra and
lightcurves obtained with a local twist applied at both polar regions produce results which, again, are not
in agreement with observations (see Fig. \ref{fig:spectra_phase2lobi}).

\section{Discussion and Conclusions}\label{sec:conclusions}

As first discussed by \cite{tlk}, the external magnetic field of a magnetar likely possesses
comparable poloidal and toroidal components. Twisted magnetospheres around ultra-magnetized neutron stars
have been shown to play a crucial role in shaping the emergent spectrum of SGRs/AXPs quiescent
emission through efficient resonant scattering of thermal photons onto the charge carriers flowing
along the field lines \citep[][]{lg06,fern,ntz1}.

In this paper we tackled the problem of constructing
sheared magnetic equilibria more general than a dipole. We have show how sheared multipolar fields of arbitrary order
can be computed by generalizing previous results by \cite{wolfson} and \cite{tlk}. In order to assess
the effects of different external field topologies on the emitted spectrum and pulse profiles we run a number
of Monte Carlo simulations, using the code of \cite{ntz1}, and compared the results to those of a sheared
dipolar field. Not surprisingly, the overall spectral shape does not change in going from a dipole to higher order
multipoles and can be always described in terms of a ``blackbody plus power-law''. There are, however, quite
substantial differences among the multipoles in the spectra viewed at different angles. These
are mainly due to the different particle distribution in the magnetosphere which is directly related to
the assumed field topology.

The case of an octupolar field has a special interest because it can be used to mimic a twist localized in a
region close to the magnetic pole(s), and hence to investigate the properties of spectra produced in locally
twisted magnetospheres. We have computed model spectra and lightcurves for the cases in which the twist is
confined to one or both polar regions (each region has semi-aperture $\theta\sim 60^\circ$), by assuming that
only the polar lobes have a non-vanishing shear while the equatorial belt is potential. Quite interestingly, a
twist confined to a single lobe is the only configuration, among those we have explored, that is able
to reproduce the main features of the high-energy ($\sim 10$--200 keV) emission observed with {\em INTEGRAL} from
the AXPs 1RXS J1708-4009 and 4U 0142+61, in particular the large variation in the pulsed fraction at different
energy bands \citep[][]{dhartog1, dhartog2}.

All magnetic equilibria we discussed in this paper are
globally twisted, axially symmetric multipolar fields. Of course, these configurations are far from being
general and, even restricting to axial symmetry, represent only a subset of the solutions of the force-free
equation. The magnetic field of a magnetar is likely to be quite complex. Modelling it in terms of single
multipolar components offers a way of gaining insight on the general properties of the magnetosphere but is far
from providing a realistic picture of these sources. A major obstacle in obtaining more complete models for
the sheared field is non-linearity of the force-free equation. Given two force-free fields, $\B_1$ and
$\B_2$, the linear combination $a\B_1+ b\B_2$ (with $a$ and $b$ two constants) is itself force-free only if
$(\rot B_1) \times \B_2 + (\rot B_2) \times \B_1=0$. This implies that a generic sheared field can not be
expressed as an expansion of sheared multipoles, or, conversely, that the superposition of twisted multipoles is
not a force-free field. An obvious case in which the previous condition is satisfied is that of potential
fields. Since sheared fields depart smoothly from potential multipoles for $p\sim p_0$, for small enough twists
a linear combination of force-free twisted multipoles (all with the same twist angle) may provide an approximate
force-free field.

Despite many efforts have been devoted to develop techniques for solving the force-free equation,
$\nabla\times\B=\alpha({\boldsymbol{x}}) \B$, no general, affordable method has been presented so far. The case
of $\alpha$ a constant has been discussed long ago by \cite{chandra57} and more recently by \cite{mastrame08},
in connection with magnetars. If $\alpha$ is a known function of position, \cite{cuper91} presented an
analytical method for solving the force-free equation also in the non-axisymmetric case. However, this is of
little use for the problem of constructing self-consistent force-free magnetospheres since prescribing $\alpha$
is tantamount to assign the currents which sustain the field, while for the case at hand the field and the
supporting currents depend on each other. A completely general, analytical technique has been proposed by
\cite{uchi97a,uchi97b}. This is based on a relativistic (tensor) description of the electromagnetic field and on
the introduction of two scalar potentials which are the analogues of the classical Euler potentials. It has been
shown to be workable in the axisymmetric case \citep[a non-aligned rotator,][]{uchi98} and, for the particular
case of a non-rotating, aligned magnetosphere, it is possible to verify that equation (\ref{ODE}) is recovered.
Further work on this is in progress and will be reported in a subsequent paper (Pavan et al., in preparation).

\section*{Acknowledgments}

The work of RT and LN is partially supported by INAF-ASI through grant AAE TH-058. SZ acknowledges STFC for
support through an Advanced Fellowship.

\appendix{}\section{} \label{sec:app}

We start by expressing both the generating function
and the eigenvalue $C$ in terms of a series expansion around
the corresponding known untwisted quantities (labelled again with the index 0)
\begin{eqnarray*}
  & &C=C_0 + \left(\frac{\ud C}{\ud p}\right)_{p_0} \Delta p + \dots\\ 
  & &f(\mu) = f_0(\mu) + f_1(\mu) \Delta p + \dots
\end{eqnarray*}
where $\Delta p\equiv p-p_0$. By substituting the previous expressions into equation~(\ref{ODE}), we obtain, to
first order in $\Delta p$,

\begin{eqnarray}
  \label{analytic}
  & &(1- \mu^2) f''_1(\mu) + p_0(1+p_0)f_1(\mu) + (1+2p_0)f_0  + \\\nonumber
  & &\displaystyle{\left(\frac{\ud C}{\ud p}\right)_{p_0}} f_0^{1+2/p_0} =0 \,  .
\end{eqnarray}

\begin{figure}
  \centering
  \includegraphics[width=0.44 \textwidth]{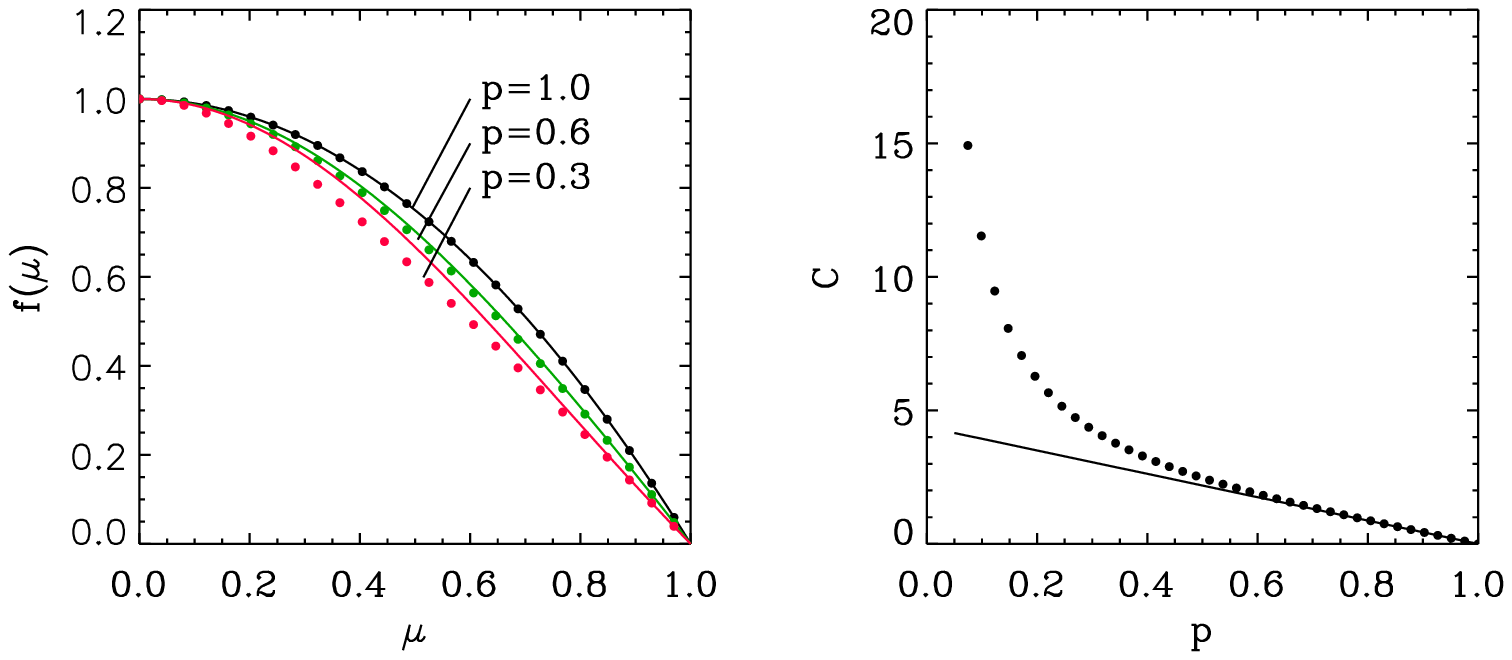}
  \includegraphics[width=0.44 \textwidth]{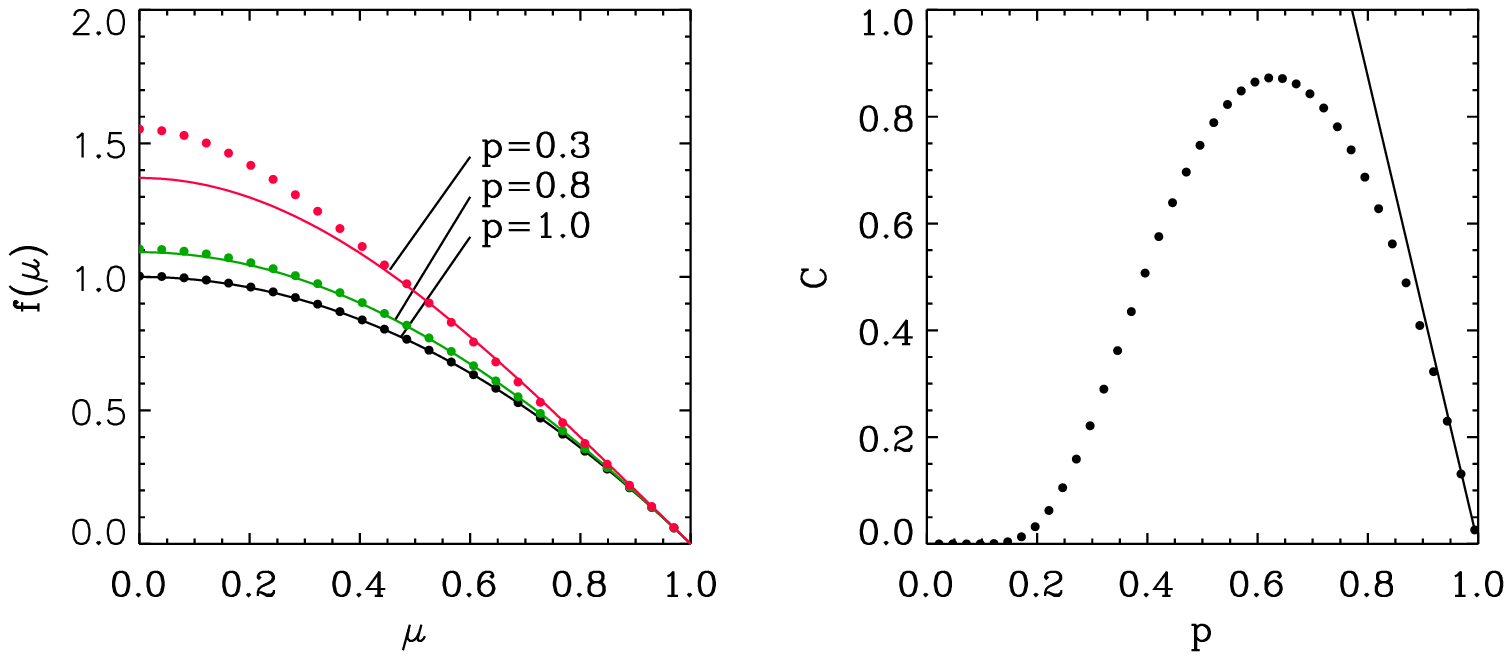}
  \caption{Dipolar angular functions $f(\mu)$ and
    eigenvalues $C(p)$
    obtained with fixed flux (upper panels)
    and fixed $B_{\textrm{pole}}$ (lower panels).
    Solid lines represent the analytical
    first order approximation, while the dots
    are the numerical solutions
    of equation~(\ref{ODE}). }
  \label{fig:1stfit}
\end{figure}

It can be easily seen that equation (\ref{analytic}) admits analytical solutions only for
integer values of the exponent $2/p_0$, i.e. for  dipoles and quadrupoles for which $2/p_0 =2$ and 1, respectively.
Since $f_0$ is itself a solution of the GSS equation, it satisfies the same boundary conditions
we need to impose on $f$. This implies that the conditions on $f_1$ are, in the case of a dipolar
field, $f_1(1) = f_1'(0)= 0$, supplemented by either $f_1'(1)=0$ or $f_1(0)=0$.
The two solutions that obey the previous two sets of conditions
are

\begin{eqnarray*}
  \label{dip1}
 & & f_1(\mu) = f_0 \frac{-17 +22\mu^2 -5\mu^4}{32} \\
  & &\left(\frac{\ud C}{\ud p}\right)_{p_0}  = -\frac{35}{8}
\end{eqnarray*}
and
\begin{eqnarray*}
  & & f_1(\mu) = f_0\frac{22\mu^2 -5\mu^4}{32}\\
  & &\left(\frac{\ud C}{\ud p}\right)_{p_0}  = -\frac{35}{8},
\end{eqnarray*}
respectively, where $f_0(\mu)= 1-\mu^2$. The complete expressions for the two generating functions are then
\begin{eqnarray*}
  & & f_{W95}\sim f_0\left(1+\frac{ 5 f_0+17}{32}\mu^2 \Delta p \right) \\
  & & f_{TLK} \sim f_{W95} - \frac{17}{32}f_0\Delta p;
\end{eqnarray*}
since $\Delta p < 0$, it is $f_{W95} < f_{TLK} $ for
any value of the parameter
in accordance with numerical results.
A comparison of the first order approximations with the exact numerical
solutions is shown in Fig.~\ref{fig:1stfit}. We find that the agreement between the two
is satisfactory (relative error $\la 8\%$) up to
$\Delta\phi_{NS} = 0.2$ in the case of fixed intensity and
$\Delta\phi_{NS} = 0.4$ in the case of constant flux.
The ratio $\lambda=f_{TLK}/f_{W95}$ introduced in \S\ref{sec:dipolar-fields} can be expanded as
\begin{equation*}
  \lambda(p)= 1- \frac{17}{32}\Delta p.
\end{equation*}

\begin{figure}
  \centering
  \includegraphics[width=0.44 \textwidth]{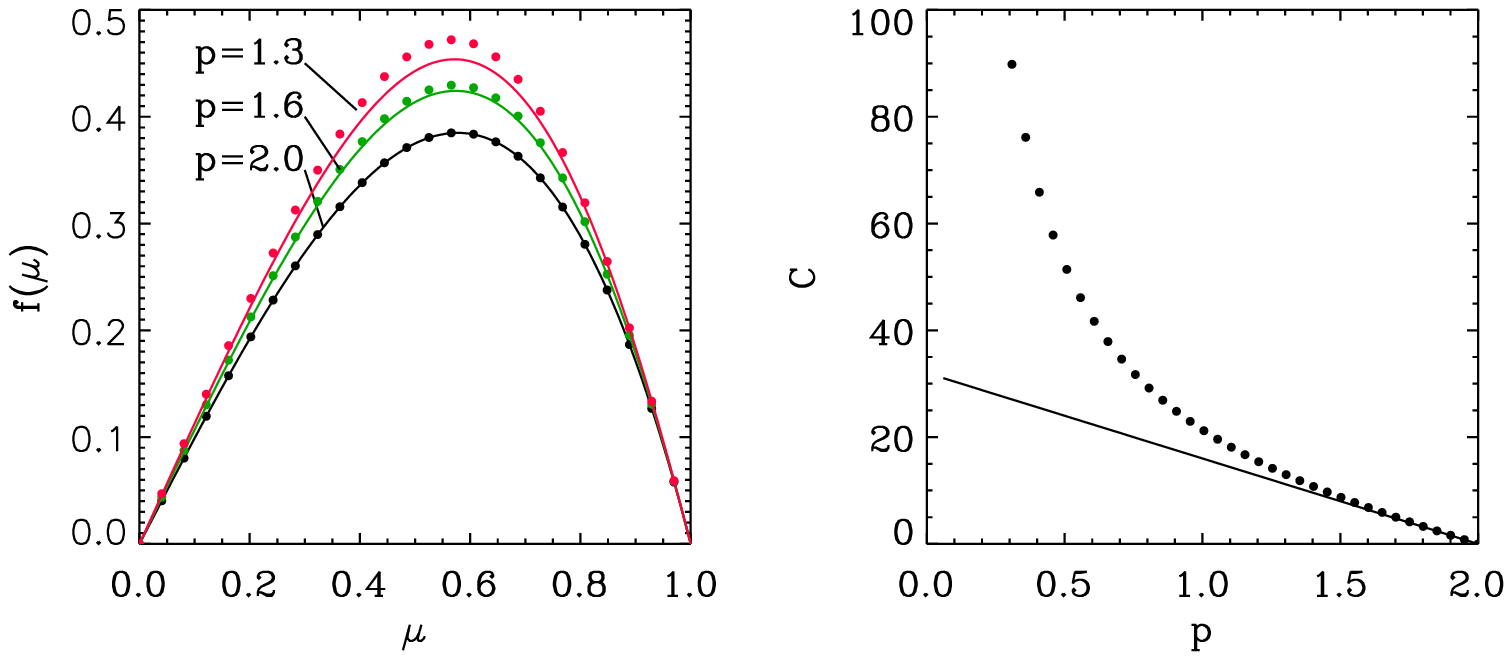}
  \caption{Same as in Fig.~\ref{fig:1stfit} for the
    quadrupolar field.}
  \label{fig:1stfit_quadr}
\end{figure}

By applying the same procedure, one can derive
the analytic first order expansion for quadrupolar fields. For
$f_1(1)=f_1(0)=f'_1(1)=0$, the first two terms in the expansion of $f$ turn out to be
\begin{eqnarray*}
  & & f_0(\mu) = \mu(1-\mu^2) \\
  & & f_1(\mu) = f_0 \left[\frac{3(1-\mu)-8\mu^2}{6(1+\mu)}
    +\frac{2}{3}\mu^3 +\ln(1+\mu)-\ln2 \right]\\
\noalign{\medskip
{\rm together with}
\medskip}
  & & \left(\frac{\ud C}{\ud p}\right)_{p_0} = -16 \, .
\end{eqnarray*}
The generating function $f$ and $C(p)$ are shown, together with the numerical solutions, in Fig. \ref{fig:1stfit_quadr}.

\label{lastpage}
\end{document}